\documentclass[11pt, a4paper]{article}
\usepackage[utf8]{inputenc}
\usepackage{authblk}
\usepackage{enumitem}
\usepackage[babel=true]{csquotes}
\usepackage{caption}
\usepackage{subcaption}
\usepackage{color}
\usepackage{dsfont}
\usepackage{booktabs}
\usepackage{capt-of}
\usepackage{multicol}

\usepackage{appendix}
\usepackage{float}
\usepackage{siunitx}
\usepackage{url}
\usepackage[colorlinks=true,bookmarks=false,linkcolor=blue,citecolor=blue,urlcolor=blue]{hyperref}
\usepackage{array,tabu,multirow,makecell}
\usepackage{threeparttable}
\usepackage{array}
\usepackage{longtable}
\usepackage{rotating}
\usepackage{tabularx}
\usepackage{lastpage}
\usepackage{lscape}
\usepackage{stackengine}
\usepackage{amsmath} 
\usepackage{amsfonts}
\usepackage[T1]{fontenc}
\usepackage{amsthm}
\usepackage{geometry}
\usepackage{graphicx}
\usepackage{amssymb}
\usepackage{bbm}
\usepackage{moreverb}
\usepackage{color, pdfcolmk}
\usepackage{fancyhdr}
\usepackage{csquotes}  
\usepackage{etoolbox}
\usepackage{rotating}
\makeatletter
\def\@footnotecolor{blue}
\define@key{Hyp}{footnotecolor}{%
 \HyColor@HyperrefColor{#1}\@footnotecolor%
}
\patchcmd{\@footnotemark}{\hyper@linkstart{link}}{\hyper@linkstart{footnote}}{}{}
\makeatother
\hypersetup{footnotecolor=blue}
\theoremstyle{plain}% default
\usepackage[round]{natbib}
\usepackage[english]{babel}
\usepackage{algorithm}
\usepackage{algorithmic}
\usepackage{pgfplots}
\pgfplotsset{compat=1.18}
\usetikzlibrary{arrows.meta,decorations.pathreplacing,positioning}

\title{Nonlinear and Heavy-Tailed Predictability in Transition-Energy Financial Markets}

\author[1]{Emmanuel GNANDI \thanks{\texttt{kpanteemmanuel@gmail.com}}
}
\author[2]{Fredy POKOU \thanks{\texttt{fredypokou@gmx.fr}}}

\author[3]{Jules SADEFO KAMDEM \thanks{\texttt{jules.sadefo-kamdem@umontpellier.fr}}}

\affil[1]{INSA de Toulouse, Département de Génie Mathématique, Toulouse, France}

\affil[2]{Inria, CNRS, Univ. of Lille, Centrale Lille, UMR 9189 - CRIStAL, F-59000 Lille, France}

\affil[2,3]{MRE UR 209 and Faculty of Economics. Montpellier University, France}

\date{\vspace{0.2cm} \small All authors contributed equally to this work. \\ \vspace{0.5cm} \today}
\begin{document}

\maketitle

\begin{abstract}

Transition-related financial markets are increasingly exposed to abrupt repricing episodes, elevated volatility, and heterogeneous macro-financial shocks. Under such conditions, conventional Gaussian-linear forecasting frameworks may provide an incomplete representation of the dependence structure linking fossil-energy, renewable-energy, technology, and utility-sector assets. This paper investigates whether transition-related financial returns exhibit residual nonlinear predictability after controlling for heavy-tailed multivariate linear dynamics. To address this question, we develop a hybrid forecasting framework combining Student-$t$ Vector Autoregressions with nonlinear recurrent residual learning architectures. The empirical analysis considers six major exchange-traded funds representing broad equity markets and key transition-sensitive sectors. The results reveal substantial departures from Gaussian-linear behavior, including excess kurtosis, volatility clustering, and remaining nonlinear dependence after econometric filtering. Out-of-sample forecasting experiments show that the proposed framework consistently improves predictive accuracy relative to conventional VAR models, standalone machine-learning methods, and alternative hybrid specifications. The forecasting gains become more pronounced during periods of macro-financial stress, particularly during the COVID-19 crisis and the Ukraine-related energy shock. Overall, the findings suggest that transition-related financial systems exhibit regime-sensitive and heavy-tailed predictive dynamics that are insufficiently captured by standard Gaussian-linear models alone.

\end{abstract}

\begin{keywords}
Transition finance; Energy-market dependence; Heavy-tailed forecasting; Student-$t$ VAR; Hybrid econometric--machine learning models; Regime-sensitive predictability
\end{keywords}

\section{Introduction}

The transition toward low-carbon economies is profoundly reshaping the dynamics of global financial markets. Over the last decade, climate-transition mechanisms have progressively altered capital allocation across fossil-energy industries, renewable-energy sectors, technology-intensive firms, and regulated utility infrastructures. These reallocations are increasingly influenced by evolving expectations regarding decarbonization policies, energy security, technological innovation, inflationary pressures, and monetary-policy normalization \citep{bolton2020green,battiston2017climate}. As a consequence, climate-transition risk has emerged as a major source of financial uncertainty with direct implications for asset pricing, portfolio rebalancing, and cross-market dependence structures.

Recent macroeconomic and geopolitical episodes have substantially amplified these dynamics. The COVID-19 collapse, the subsequent inflationary tightening cycle, and the Ukraine-related energy shock simultaneously affected fossil-energy markets, renewable-energy assets, technology sectors, and broad equity indices. Importantly, these episodes revealed that transition-related financial adjustments do not evolve through smooth or stable repricing mechanisms. Instead, periods of macro-financial stress are frequently associated with abrupt volatility amplification, persistent dislocations, and highly nonlinear market reactions across interconnected financial sectors.

An expanding literature has therefore investigated the financial implications of climate-transition dynamics. Existing studies document significant interactions between oil prices, renewable-energy assets, technology sectors, and carbon-related financial variables \citep{henriques2008oil,sadorsky2012correlations,kumar2012stock}. More recent contributions further emphasize the systemic implications of transition-related uncertainty for financial stability and asset valuation \citep{battiston2017climate,bolton2020green,bolton2021investors,ilhan2021carbon}. In parallel, the connectedness literature shows that periods of elevated uncertainty tend to amplify cross-market dependence and destabilize conventional diversification structures \citep{diebold2012better}.\\

Despite these advances, the econometric characterization of transition-energy financial systems remains incomplete. In particular, two empirical regularities appear especially difficult to reconcile within conventional forecasting frameworks.\\

First, transition-sensitive financial assets exhibit pronounced heavy-tailed behavior. Renewable-energy and fossil-energy markets are repeatedly exposed to abrupt repricing episodes associated with geopolitical disruptions, inflationary shocks, energy-supply uncertainty, and revisions in climate-policy expectations. Such episodes generate excess kurtosis, volatility clustering, and extreme return realizations that are inconsistent with Gaussian innovations.\\

Second, the predictability structure of transition-related financial markets appears strongly regime-dependent. During periods of macro-financial stress, the interaction between fossil-energy repricing, renewable-energy expectations, technology-sector valuations, and macroeconomic uncertainty generates dependence structures that evolve asymmetrically across market conditions. Relatively localized shocks may propagate simultaneously through financing conditions, inflation expectations, portfolio reallocations, and energy-price adjustments, thereby amplifying nonlinear dependence across sectors.\\

These empirical characteristics create important limitations for standard forecasting approaches. Conventional linear Vector Autoregressive (VAR) systems rely on Gaussian innovations and relatively stable dependence structures that may underestimate transition-related tail risk and nonlinear predictability during turbulent periods. Conversely, purely machine learning approaches often capture nonlinear patterns effectively but generally lack explicit multivariate dependence structures and economically interpretable forecasting mechanisms.
Consequently, the existing literature frequently addresses either heavy-tailed dynamics or nonlinear forecasting separately, while rarely integrating both dimensions within a unified framework designed to characterize regime-sensitive predictability in transition-energy financial markets.\\

This paper investigates whether the predictability structure of transition-energy financial markets changes systematically across macro-financial stress regimes. To address this question, we propose a hybrid forecasting architecture combining Student-\textit{t} Vector Autoregressions with nonlinear recurrent residual learning. The central idea is economically intuitive. The econometric component captures heavy-tailed cross-market dependence and interpretable linear interactions across transition-related financial assets, while the recurrent nonlinear component models the remaining sequential dependence embedded within the residual dynamics.

The empirical analysis focuses on six major exchange-traded funds representing complementary dimensions of transition-related financial markets:
\begin{itemize}
    \item broad equity markets (SPY),
    \item technology-intensive growth sectors (QQQ),
    \item fossil-energy industries (XLE),
    \item renewable-energy markets (ICLN and TAN),
    \item and utility infrastructure assets (XLU).
\end{itemize}

Taken together, these assets provide an empirical representation of the broader transition-finance ecosystem linking decarbonization dynamics, energy-market repricing, technological expectations, and macro-financial adjustment.\\

\noindent The contribution of the paper is threefold.

First, the paper documents that transition-energy financial markets exhibit substantial heavy-tailed behavior, volatility persistence, and nonlinear residual dependence that intensify during macro-financial stress regimes. In particular, the empirical evidence reveals that the predictability structure of transition-related assets changes significantly during periods such as the COVID-19 collapse, the post-pandemic recovery, the inflation-tightening episode, and the Ukraine-related energy shock.

Second, the paper develops a hybrid heavy-tailed forecasting framework combining Student-\textit{t} VAR models with recurrent residual-learning architectures. Methodologically, the proposed specification jointly captures:
\begin{enumerate}
    \item heavy-tailed return innovations,
    \item multivariate cross-market dependence,
    \item and nonlinear sequential predictability.
\end{enumerate}

Unlike purely black-box forecasting systems, the proposed architecture preserves an interpretable econometric structure while allowing flexible nonlinear correction of residual dynamics.

Third, the paper provides extensive out-of-sample evidence showing that hybrid heavy-tailed architectures substantially outperform both conventional econometric specifications and standalone machine learning systems across all transition-related asset classes. More importantly, forecasting gains become considerably larger during crisis and stress regimes, suggesting that nonlinear and heavy-tailed predictability intensifies precisely when macro-financial uncertainty increases.

Overall, the findings indicate that transition-energy financial markets cannot be adequately characterized through Gaussian-linear forecasting structures alone. Instead, the empirical evidence supports the existence of regime-sensitive nonlinear and heavy-tailed predictability mechanisms linking macro-financial stress, energy-market repricing, and transition-related financial dynamics.

\section{Related Literature}

This paper relates to four interconnected strands of literature: climate-transition finance, financial connectedness in energy markets, heavy-tailed financial econometrics, and hybrid econometric--machine learning forecasting. Although these research areas have developed substantially over the last two decades, the existing literature remains fragmented both conceptually and methodologically.

More specifically, most existing contributions focus on only one dimension of transition-related financial dynamics at a time. The climate-finance literature primarily emphasizes transition risk and carbon-related repricing mechanisms, often within static or linear frameworks. The connectedness literature studies cross-market dependence and volatility transmission but generally relies on Gaussian-linear specifications. The heavy-tailed econometric literature accounts for excess kurtosis and tail risk, yet remains only weakly integrated with multivariate transition-finance forecasting. Finally, the hybrid econometric--machine learning literature improves predictive performance but often sacrifices economic interpretability for forecasting flexibility.

As a consequence, relatively few studies simultaneously address:
\begin{enumerate}
    \item heavy-tailed return dynamics,
    \item nonlinear forecasting behavior,
    \item regime-sensitive predictability,
    \item and multivariate dependence structures
\end{enumerate}
within a unified transition-finance framework.

The present paper contributes precisely along these dimensions by developing a hybrid Student-\textit{t} VAR and recurrent residual-learning framework designed for regime-sensitive forecasting in transition-energy financial markets.
Table~\ref{tab:literature_positioning} summarizes the positioning of the present paper relative to the existing literature.

\begin{table}[ht]
\centering
\caption{Positioning of the Present Paper within the Existing Literature}
\label{tab:literature_positioning}
\resizebox{18cm}{!}{
\begin{tabular}{p{3cm} p{3cm} c c c c p{4cm}}
\toprule
\textbf{Study} &
\textbf{Main Focus} &
\textbf{Heavy-Tailed Dynamics} &
\textbf{Nonlinear Forecasting} &
\textbf{Regime Sensitivity} &
\textbf{Multivariate Structure} &
\textbf{Main Limitation} \\
\midrule

\cite{henriques2008oil}
& Oil and alternative-energy interactions
& No
& No
& No
& Partial
& Static linear dependence \\

\cite{sadorsky2012correlations}
& Oil-clean energy volatility interactions
& Partial
& No
& Partial
& Yes
& Gaussian dependence structure \\

\cite{diebold2012better}
& Financial connectedness
& No
& No
& Partial
& Yes
& Linear dependence assumption \\

\cite{battiston2017climate}
& Climate-related systemic risk
& Partial
& No
& Yes
& Yes
& No forecasting perspective \\

\cite{ilhan2021carbon}
& Carbon tail risk
& Yes
& No
& Partial
& Partial
& Limited dynamic forecasting structure \\

\cite{pokou2024hybridization}
& Hybrid heavy-tailed forecasting
& Yes
& Yes
& No
& No
& Univariate forecasting framework \\

\cite{pokou2026predictive}
& Energy-macro forecasting
& Yes
& Yes
& Yes
& Yes
& No residual recurrent learning \\

\cite{gnandi2026hybrid}
& Hybrid multivariate forecasting
& Partial
& Yes
& Partial
& Yes
& No explicit heavy-tailed specification \\

\textbf{Present paper}
& Regime-sensitive forecasting in transition-energy markets
& Yes
& Yes
& Yes
& Yes
& Hybrid heavy-tailed residual learning \\

\bottomrule
\end{tabular}
}
\end{table}

\subsection{Climate Transition Finance and Energy-Market Interactions}

Early contributions to the energy-finance literature primarily focused on the interaction between oil prices and alternative-energy assets. These studies emerged from the observation that renewable-energy firms and fossil-energy markets are intrinsically connected through substitution effects, technological competition, and energy-demand expectations.

A seminal contribution is provided by \citet{henriques2008oil}, who analyze the relationship between oil prices and alternative-energy stock prices. Their findings suggest that oil-market fluctuations significantly affect renewable-energy equity valuation, thereby highlighting the financial interdependence between traditional and alternative-energy sectors. Similarly, \citet{sadorsky2012correlations} document important volatility interactions between oil prices, clean-energy companies, and technology stocks, emphasizing the growing integration of renewable-energy assets within broader financial markets.

Subsequent research progressively evolved toward the broader notion of climate-transition finance. Rather than focusing exclusively on oil-price dynamics, this literature investigates how decarbonization policies, carbon regulation, climate uncertainty, and green technological innovation reshape financial markets and asset-pricing mechanisms.

An important milestone was introduced by \citet{battiston2017climate}, who develop a climate stress-testing framework for financial systems. Their analysis shows that transition-related shocks may propagate across financial networks through interconnected exposures and contagion channels. Likewise, \citet{bolton2020green} argue that climate change constitutes a systemic source of financial instability capable of generating abrupt repricing episodes, stranded-asset risk, and large-scale portfolio reallocations.

A rapidly expanding branch of the literature further investigates whether investors explicitly price transition-related risk. Using firm-level emissions data, \citet{bolton2021investors} show that firms with higher carbon emissions command higher expected returns, suggesting that climate-transition risk is increasingly incorporated into financial valuation. Similarly, \citet{ilhan2021carbon} introduce the notion of carbon tail risk and demonstrate that climate uncertainty generates substantial downside-tail exposure in financial markets.

Collectively, these studies establish that transition-energy financial systems are characterized by unstable dependence structures, asymmetric responses to shocks, and heightened sensitivity to macro-financial stress. However, most existing contributions continue to rely on linear or Gaussian specifications that may inadequately capture the nonlinear and heavy-tailed dynamics frequently observed during turbulent periods.

\subsection{Financial Connectedness and Regime-Dependent Dependence Structures}

The analysis of financial connectedness has become a central theme in modern financial econometrics. The foundational contribution of \citet{diebold2012better} introduced directional measures of volatility spillovers based on forecast-error variance decompositions within Vector Autoregressive systems. Their framework demonstrated that systemic risk transmission and financial contagion can be quantified through dynamic connectedness measures.

This methodology rapidly became influential in energy and commodity-market research. Subsequent studies document that energy markets exhibit substantial dependence dynamics, particularly during periods of geopolitical instability, commodity-price shocks, and macroeconomic stress. Renewable-energy assets also appear increasingly integrated with broader equity and technology markets, thereby amplifying cross-market dependence during turbulent episodes.

Nevertheless, despite the sophistication of connectedness measures, most existing frameworks remain fundamentally linear and Gaussian. Standard VAR specifications implicitly assume symmetric dependence structures and finite-variance innovations, assumptions that may become problematic in transition-related environments characterized by abrupt repricing episodes and extreme return realizations.

To overcome these limitations, recent contributions increasingly incorporate nonlinear and time-varying dependence structures through copula-based models, DCC-GARCH specifications, and Time-Varying Parameter VAR systems. In particular, \citet{pokou2026predictive} compare copula-enhanced TVP-SVAR systems with advanced machine learning approaches for forecasting energy and macroeconomic variables. Their results show that structurally interpretable econometric systems may achieve forecasting performances comparable to machine learning while preserving economically meaningful regime diagnostics.

However, an important limitation remains. Most existing connectedness models either prioritize interpretability or predictive flexibility separately. Relatively few studies explicitly investigate how nonlinear predictability and heavy-tailed dependence evolve across macro-financial regimes within transition-energy financial systems.

\subsection{Heavy-Tailed Financial Econometrics}

The inadequacy of Gaussian assumptions for financial returns has long been recognized in econometric theory. Beginning with the ARCH and GARCH literature, researchers progressively documented that financial returns exhibit excess kurtosis, volatility clustering, and leptokurtic distributions inconsistent with Gaussian innovations.

In particular, \citet{bollerslev1987conditionally} demonstrate that Student-\textit{t} conditional distributions substantially improve volatility modeling in financial markets. Likewise, \citet{baillie2002message} emphasize the importance of heavy-tailed innovations when analyzing speculative-price dynamics and exchange-rate behavior.

Heavy-tailed econometric models became especially relevant in commodity and energy markets due to the prevalence of geopolitical disruptions, oil-price shocks, macroeconomic crises, and climate-policy uncertainty. Such environments frequently generate abrupt tail events and extreme market movements that cannot be adequately represented through Gaussian innovations.

More recently, the climate-finance literature increasingly emphasizes that renewable-energy and transition-related assets exhibit stronger downside-tail exposure and greater sensitivity to systemic stress than traditional equity sectors. The emergence of climate-transition risk therefore reinforces the importance of heavy-tailed econometric specifications in transition-related financial modeling.

Despite these advances, the heavy-tailed econometric literature remains largely disconnected from nonlinear multivariate forecasting frameworks. In particular, relatively few studies integrate heavy-tailed innovations with residual nonlinear learning within transition-finance systems characterized by regime-sensitive predictability.

\subsection{Hybrid Econometric--Machine Learning Forecasting}

The rapid development of machine learning techniques has profoundly transformed forecasting research in finance and macroeconomics. Methods such as Support Vector Regression, Artificial Neural Networks, Long Short-Term Memory networks, and Gated Recurrent Units are increasingly used to model nonlinear temporal dependence structures.

However, purely machine learning approaches often suffer from important limitations in financial applications, including limited interpretability, instability across regimes, and overfitting risk. As a consequence, a growing literature has developed around hybrid econometric--machine learning systems combining the structural interpretability of econometric models with the nonlinear approximation capabilities of machine learning algorithms.

Early hybrid systems primarily relied on ARIMA-based decompositions in which econometric models capture linear dependence structures while machine learning methods estimate nonlinear residual components. More recently, these approaches have been extended to multivariate forecasting environments. For example, \citet{gnandi2026hybrid} propose hybrid VARIMA--machine learning architectures for macroeconomic and energy forecasting, showing that residual-learning systems may substantially improve predictive performance during periods of structural instability.

Similarly, \citet{pokou2024hybridization} develop hybrid ARIMA--machine learning systems with Student-\textit{t} innovations for financial forecasting. Their findings indicate that introducing heavy-tailed disturbances substantially improves the ability of hybrid systems to capture leptokurtic financial dynamics and crisis episodes.

Despite these advances, several limitations remain within the hybrid forecasting literature. First, many existing hybrid systems remain primarily prediction-oriented and provide limited economic interpretation of the underlying forecasting mechanisms. Second, most studies continue to rely implicitly on Gaussian assumptions despite overwhelming evidence of heavy-tailed financial returns. Third, relatively little attention has been devoted to regime-sensitive predictability in transition-energy financial systems.

The present paper contributes precisely along these dimensions by proposing a hybrid Student-\textit{t} VAR and recurrent residual-learning framework designed to capture heavy-tailed and nonlinear predictability across distinct macro-financial regimes. Unlike purely black-box forecasting architectures, the proposed framework preserves an interpretable multivariate econometric structure while allowing flexible nonlinear residual correction during periods of elevated market stress.

\section{Economic Rationale and Empirical Setting}

The transition toward low-carbon economies is generating profound transformations in the organization of financial markets. Beyond a purely technological adjustment, the transition process involves large-scale reallocations of capital across fossil-energy industries, renewable-energy sectors, technological innovation systems, and regulated utility infrastructures \citep{bolton2020green,battiston2017climate}. These reallocations are increasingly shaped by climate-policy expectations, energy-security concerns, inflationary pressures, financing conditions, and changing expectations regarding long-run decarbonization trajectories.

From a financial perspective, the transition process creates a highly interconnected environment in which asset valuations become increasingly sensitive to both macroeconomic conditions and transition-specific shocks. Energy-price fluctuations, monetary-policy adjustments, geopolitical disruptions, and climate-policy announcements may simultaneously affect fossil-energy firms, renewable-energy companies, technology-intensive sectors, and utility infrastructures. 

\newpage

As a consequence, transition-related financial systems are characterized by evolving dependence structures, episodic volatility amplification, and potentially unstable forecasting dynamics.
Rather than interpreting transition-related assets as isolated financial instruments, this paper considers them as interconnected components of a broader transition-energy financial system. Exchange-traded funds (ETFs) provide particularly useful empirical representations of this system because they aggregate market expectations and facilitate rapid portfolio reallocation across sectors. Their dynamics therefore reflect the interaction between macroeconomic conditions, technological innovation, energy-market repricing, and investor expectations regarding the transition process.\\

The empirical framework relies on six major ETFs representing complementary dimensions of transition-related financial markets:
\begin{itemize}
    \item aggregate equity-market conditions,
    \item technology-intensive growth sectors,
    \item fossil-energy exposure,
    \item renewable-energy development,
    \item solar-energy specialization,
    \item and utility infrastructure adaptation.
\end{itemize}
Taken together, these assets provide a parsimonious but economically coherent representation of the transition-energy financial ecosystem.
Importantly, the objective of this section is not to claim the existence of deterministic structural contagion mechanisms or fully identified causal transmission channels. Instead, the goal is to provide an economically grounded interpretation of the main cross-market interactions that motivate the multivariate forecasting framework developed in the remainder of the paper.

\subsection{Fossil and Renewable Energy Reallocation Dynamics}

One of the central dimensions of the transition process concerns the progressive reallocation of financial capital between fossil-energy industries and renewable-energy sectors. This reallocation is influenced by multiple factors, including environmental regulation, technological progress, investor preferences, energy-demand expectations, and geopolitical conditions \citep{battiston2017climate,bolton2021investors}.
Within the present framework, XLE proxies the traditional fossil-energy sector, whose valuation remains closely connected to hydrocarbon prices, global energy demand, and geopolitical developments. By contrast, ICLN and TAN capture renewable and solar-energy markets whose profitability depends more strongly on renewable deployment, subsidy regimes, financing conditions, and expectations regarding future decarbonization policies.\\

The interaction between these sectors does not necessarily imply stable or monotonic substitution effects. Periods of rising oil prices may temporarily improve the relative attractiveness of fossil-energy firms while simultaneously increasing incentives for renewable-energy investment. Conversely, tightening monetary conditions may disproportionately affect renewable-energy firms because of their sensitivity to long-duration financing costs and growth expectations.\\

The existing literature documents substantial interactions between oil markets, renewable-energy assets, and technology-intensive sectors. For example, \citet{henriques2008oil} and \citet{sadorsky2012correlations} show that oil-price fluctuations significantly affect renewable-energy equity dynamics and broader financial markets. More recent studies further emphasize the growing integration between climate-transition expectations and financial-market behavior \citep{bolton2020green,ilhan2021carbon}.

From an econometric perspective, these interactions suggest that transition-related assets may exhibit evolving cross-market dependence structures, particularly during periods of elevated uncertainty and abrupt repricing.

\subsection{Technology and Innovation-Sensitive Financial Dynamics}

Technological innovation constitutes another important component of the transition process. Renewable-energy deployment, electrification systems, battery storage technologies, and digital energy-management infrastructures all rely heavily on innovation-intensive sectors and long-duration investment dynamics.

In the empirical framework, QQQ serves as a proxy for technology-intensive growth sectors that are closely connected to innovation cycles and future-oriented investment expectations. Technology firms increasingly participate in renewable-energy financing, energy-efficiency systems, AI-based optimization technologies, and infrastructure digitalization.

Importantly, technology-intensive and renewable-energy assets share several common financial characteristics. Both sectors are highly sensitive to:
\begin{itemize}
    \item interest-rate fluctuations,
    \item financing conditions,
    \item long-duration cash-flow expectations,
    \item innovation cycles,
    \item and investor sentiment.
\end{itemize}
This common sensitivity may amplify dependence structures between renewable-energy and technology-related assets during periods of macro-financial stress. For example, tightening monetary conditions may simultaneously affect clean-energy and technology sectors through discount-rate adjustments and deteriorating financing conditions. Conversely, favorable climate-policy announcements may stimulate both renewable-energy valuations and technology-intensive growth expectations.
Such interactions do not necessarily imply deterministic spillover mechanisms. Rather, they suggest that the predictability structure of these assets may evolve jointly across macro-financial regimes.

\subsection{Utilities and Infrastructure Stabilization}

Utilities occupy a distinct position within the transition-energy financial system. Unlike renewable-energy sectors, utility firms generally operate within regulated environments characterized by more stable cash flows, lower speculative exposure, and comparatively defensive market behavior.

Within the present framework, XLU proxies the infrastructure dimension of the transition process. Utility firms play a central role in integrating renewable-energy production into electricity grids, expanding transmission capacity, and supporting large-scale electrification dynamics.

At the same time, utilities face substantial long-run investment requirements associated with:
\begin{itemize}
    \item grid modernization,
    \item renewable-energy integration,
    \item energy-storage systems,
    \item and transmission-network expansion.
\end{itemize}
Consequently, utilities simultaneously exhibit defensive short-run characteristics and long-run exposure to transition-related investment cycles. This intermediate position may generate dependence structures that differ from those observed in fossil-energy and renewable-energy sectors.
From a forecasting perspective, the comparatively defensive profile of utilities may also contribute to heterogeneous dependence patterns across assets and regimes.

\subsection{Macroeconomic Stress and Regime-Sensitive Dependence}

An important feature of transition-related financial systems is that cross-market dependence structures may evolve substantially across macro-financial environments. Inflationary episodes, geopolitical disruptions, commodity-price shocks, and monetary-policy tightening may simultaneously affect financing conditions, discount rates, investor expectations, and sectoral profitability.
\newpage
\noindent In this context, periods of macro-financial stress may amplify:
\begin{itemize}
    \item volatility persistence,
    \item cross-market dependence,
    \item abrupt repricing episodes,
    \item and nonlinear forecasting dynamics.
\end{itemize}

Importantly, the objective is not to argue that all transition-related interactions are inherently nonlinear or unstable at every point in time. Instead, the central hypothesis of this paper is more modest and empirically testable: the predictability structure of transition-energy financial markets may become increasingly nonlinear and heavy-tailed during periods of elevated macro-financial uncertainty.
This interpretation directly motivates the empirical analysis developed in the following section. If transition-related financial systems simultaneously exhibit heavy-tailed return distributions, volatility clustering, and residual nonlinear dependence, then purely Gaussian-linear forecasting frameworks may become insufficient during stress regimes.

The next section investigates these empirical regularities and provides the statistical motivation for the hybrid heavy-tailed forecasting architecture proposed in the paper.

\section{Stylized Facts of Transition-Energy Financial Markets}
\label{sec4}

This section documents the principal empirical characteristics of the transition-related financial assets considered in the analysis and motivates the econometric architecture developed in the remainder of the paper. The objective is not to establish structural causality or fully identified transmission mechanisms, but rather to characterize the statistical regularities that are most relevant for forecasting transition-energy financial markets.
More specifically, the empirical evidence reveals three main stylized facts. First, return distributions exhibit substantial departures from Gaussianity characterized by heavy tails and asymmetric extreme realizations. Second, volatility dynamics display strong persistence and conditional heteroskedasticity. Third, even after filtering linear cross-market dependence through a Vector Autoregressive specification, significant residual nonlinear dependence remains.

Taken together, these findings provide the econometric justification for combining heavy-tailed multivariate econometric models with nonlinear residual-learning architectures.

\subsection{Financial Assets and Empirical Representation of the Transition System}

The empirical analysis relies on six major exchange-traded funds representing complementary dimensions of the transition-energy financial system. Table~\ref{tab:assets} summarizes the assets used throughout the study.

\begin{table}[!h]
\centering
\caption{Description of Financial Assets Used in the Empirical Analysis}
\label{tab:assets}
\resizebox{12.5cm}{!}{
\begin{tabular}{|l|lll|}
\hline
\textbf{Symbol} & \textbf{Asset Name} & \textbf{Category} & \textbf{Market Role} \\
\hline
SPY  & SPDR S\&P 500 ETF Trust                & Equity Market      & Broad US Market Benchmark \\
QQQ  & Invesco QQQ Trust                      & Equity Growth      & Technology-Heavy Index Proxy \\
XLE  & Energy Select Sector SPDR Fund         & Fossil Energy      & Fossil Fuel Sector Exposure \\
ICLN & iShares Global Clean Energy ETF        & Renewable Energy   & Global Clean Energy Exposure \\
TAN  & Invesco Solar ETF                      & Solar Energy       & Solar Industry Proxy \\
XLU  & Utilities Select Sector SPDR Fund      & Utilities Sector   & Energy Infrastructure \\
\hline
\end{tabular}
}
\end{table}

The selected assets jointly capture several economically important dimensions of transition-related financial markets. SPY proxies aggregate macroeconomic and equity-market conditions, while QQQ reflects technology-intensive growth sectors closely connected to innovation and financing conditions. XLE captures traditional fossil-energy exposure, whereas ICLN and TAN represent renewable and solar-energy sectors that are particularly sensitive to climate-policy expectations and long-duration investment dynamics. Finally, XLU proxies utility infrastructures and regulated energy systems.

This classification provides a coherent empirical representation of the transition-energy financial environment while remaining sufficiently parsimonious for multivariate forecasting analysis.

\subsection{Distributional Non-Normality and Heavy-Tailed Dynamics}

\begin{figure}[!h]
\centering
\includegraphics[width=0.8\textwidth]{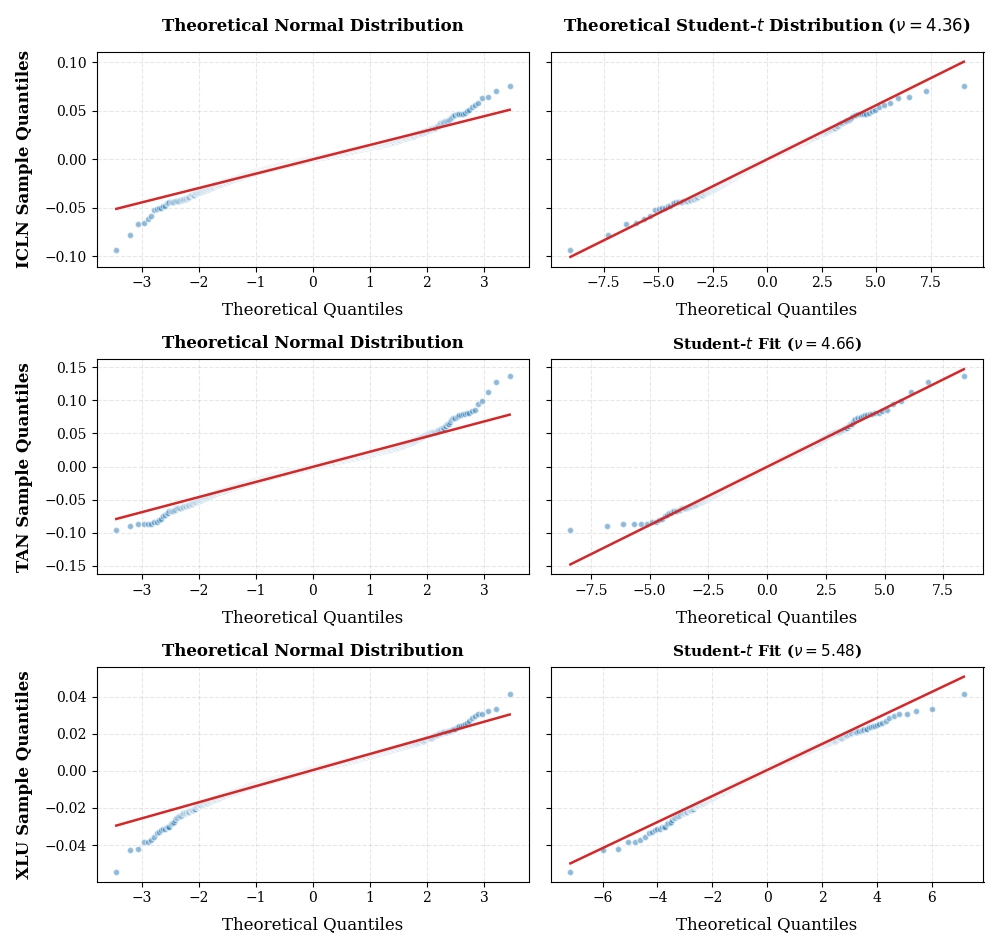}
\caption{QQ-plots comparing empirical return distributions with Gaussian and Student-\textit{t} theoretical distributions for selected assets.}
\label{fig:qqplots}
\end{figure}

Table~\ref{tab:returns_descriptive_stats} reports the descriptive statistics and diagnostic tests for the return series.
Several important features emerge from the data. First, all return series are stationary according to the Augmented Dickey-Fuller test, thereby supporting the use of VAR-based specifications. Second, all Jarque-Bera statistics strongly reject the null hypothesis of Gaussianity, indicating substantial departures from normal distributions.
The return distributions also exhibit asymmetric realizations and elevated tail risk. SPY displays a kurtosis level exceeding 4.4, while renewable-energy assets such as ICLN and TAN exhibit comparatively large extreme realizations. These findings suggest that transition-related financial markets are exposed to episodic tail events that may not be adequately represented through Gaussian innovations.
To further investigate distributional properties, Figure~\ref{fig:qqplots} compares empirical quantiles against Gaussian and Student-\textit{t} theoretical distributions.

\begin{table}[!h]
\centering
\caption{Descriptive statistics and diagnostic tests for the asset return distributions. Asymptotic significance for the Augmented Dickey-Fuller (ADF) stationarity test and the Jarque-Bera (JB) normality test is indicated by asterisks.}
\label{tab:returns_descriptive_stats}
\resizebox{13.5cm}{!}{
\begin{tabular}{@{} l rrrrrr r r @{}}
\toprule
\textbf{Asset } & \textbf{Mean} & \textbf{Std. Dev.} & \textbf{Skewness} & \textbf{Kurtosis} & \textbf{Min} & \textbf{Max} & \textbf{ADF} & \textbf{Jarque-Bera} \\
\midrule
ICLN & -0.0002 & 0.0150 & -0.2574 & 2.6291 & -0.0931 & 0.0751 & -29.1929$^{***}$ & 747.9484$^{***}$ \\
QQQ &  0.0006 & 0.0109 & -0.3922 & 3.0935 & -0.0621 & 0.0606 & -19.6823$^{***}$ & 1061.3662$^{***}$ \\
SPY &  0.0005 & 0.0093 & -0.5086 & 4.4933 & -0.0673 & 0.0493 & -11.3764$^{***}$ & 2211.7613$^{***}$ \\
TAN & -0.0004 & 0.0231 &  0.0994 & 2.4405 & -0.0960 & 0.1363 & -46.6191$^{***}$ & 624.7622$^{***}$ \\
XLE &  0.0001 & 0.0137 & -0.2853 & 2.6362 & -0.0889 & 0.0604 & -24.2084$^{***}$ & 758.1289$^{***}$ \\
XLU &  0.0004 & 0.0088 & -0.4629 & 2.3119 & -0.0545 & 0.0413 & -30.6575$^{***}$ & 646.3160$^{***}$ \\
\bottomrule
\multicolumn{9}{l}{\footnotesize \textit{Note:} $^{***}$ denotes statistical significance at the 1\% level.} \\
\end{tabular}
}
\end{table}

\newpage
The Gaussian specification systematically underestimates the magnitude of extreme observations, particularly for renewable-energy assets such as ICLN and TAN. By contrast, the Student-\textit{t} distribution provides a substantially closer approximation to the empirical tails. The estimated degrees of freedom remain relatively low, suggesting economically meaningful heavy-tailed exposure.
These findings motivate the use of heavy-tailed econometric specifications in the forecasting framework developed later in the paper.

\subsection{Volatility Persistence and Conditional Heteroskedasticity}

\begin{figure}[!h]
\centering
\includegraphics[width=1.05\textwidth]{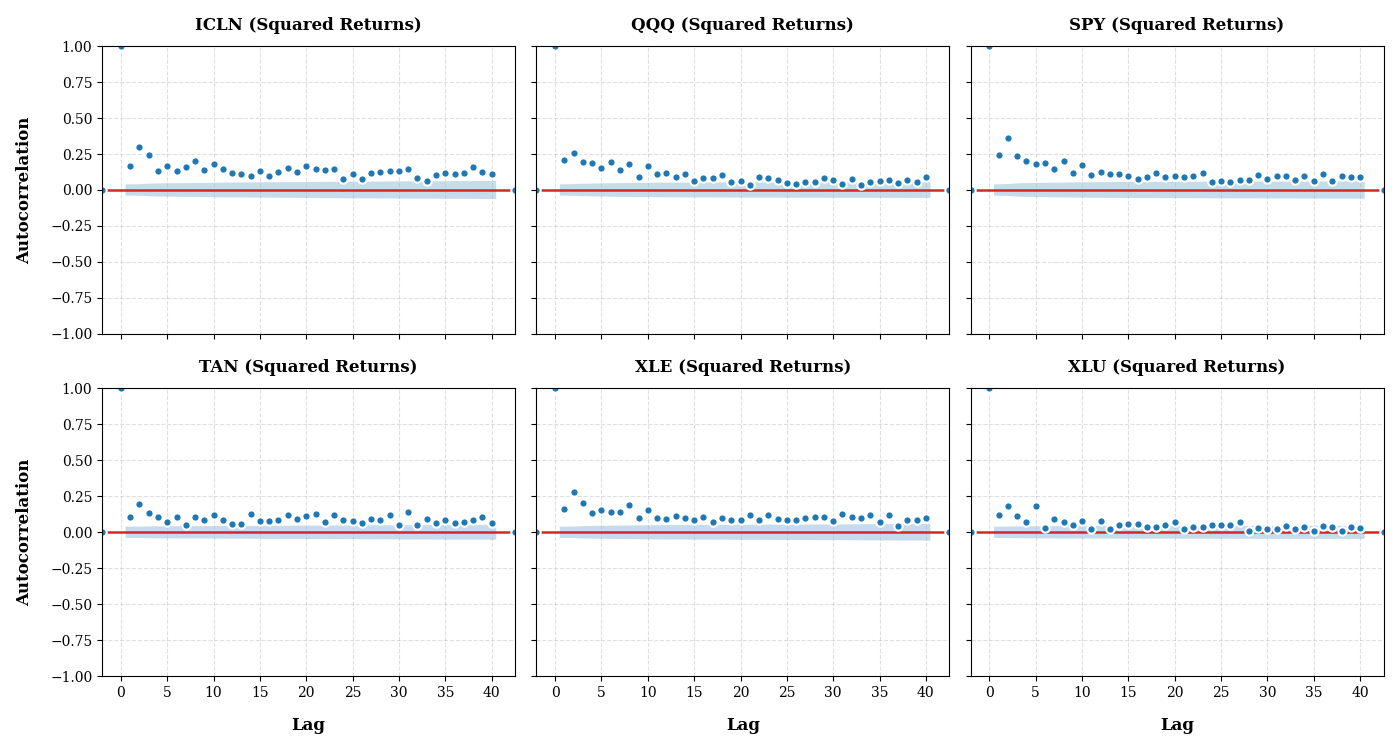}
\caption{Autocorrelation functions of squared returns for all assets.}
\label{fig:acf_squared_returns}
\end{figure}

Another important characteristic of transition-related financial assets concerns volatility persistence. Figure~\ref{fig:acf_squared_returns} reports the autocorrelation functions of squared returns for all assets.

The autocorrelation coefficients remain significantly positive across multiple lags, indicating persistent volatility clustering. Such persistence is particularly pronounced for ICLN, SPY, and XLE, suggesting prolonged adjustment phases following large shocks.
To formally assess conditional heteroskedasticity, Table~\ref{tab:arch_lm_test_results} reports Engle's ARCH-LM statistics.

\begin{table}[!h]
\centering
\caption{Engle's Lagrange Multiplier (ARCH-LM) test results for conditional heteroskedasticity. The table reports both the LM test statistic and the robust F-statistic alternative across forty lags. Significant statistics reject the null hypothesis of no ARCH effects.}
\label{tab:arch_lm_test_results}
\resizebox{9.5cm}{!}{
\begin{tabular}{@{} l cc cc @{}}
\toprule
\textbf{Asset} & \textbf{ARCH-LM Statistic} & \textbf{$p$-value} & \textbf{$F$ Statistic} & \textbf{$F$ $p$-value} \\
\midrule
ICLN & 394.5154 & $^{***}$ & 46.6685 & $^{***}$ \\
QQQ  & 336.7936 & $^{***}$ & 38.7729 & $^{***}$ \\
SPY  & 456.0423 & $^{***}$ & 55.5778 & $^{***}$ \\
TAN  & 185.6964 & $^{***}$ & 19.9768 & $^{***}$ \\
XLE  & 330.9728 & $^{***}$ & 38.0001 & $^{***}$ \\
XLU  & 187.2950 & $^{***}$ & 20.1628 & $^{***}$ \\
\bottomrule
\multicolumn{5}{l}{\footnotesize \textit{Note:} $^{***}$ indicates asymptotic statistical significance at the 0.1\% level ($p < 0.001$).} \\
\end{tabular}
}
\end{table}

\noindent The null hypothesis of homoskedasticity is overwhelmingly rejected for all assets. These findings confirm the presence of strong conditional variance dynamics throughout the transition-related financial system.
From a forecasting perspective, such persistence suggests that volatility regimes may evolve slowly over time and may interact with broader macro-financial conditions.
\newpage
\subsection{Cross-Market Dependence Structure}

To characterize multivariate dependence dynamics, we estimate a Vector Autoregressive model. Table~\ref{tab:var_lag_selection} reports the lag-order selection criteria.
The Akaike Information Criterion selects a parsimonious VAR(1) specification, suggesting that an important fraction of short-run linear dependence can be captured through low-order multivariate dynamics.
Table~\ref{tab:residual_correlation} reports the contemporaneous residual correlation matrix obtained from the VAR(1) specification.
Several economically plausible dependence patterns emerge. Renewable-energy assets exhibit strong contemporaneous association, while SPY and QQQ display very high correlation consistent with common macroeconomic and technology-related factors. Utilities exhibit comparatively weaker dependence with the remaining assets, consistent with their more defensive profile.

\begin{table}[!h]
\centering
\caption{Vector Autoregression (VAR) lag order selection criteria. Minimum values for each information criterion are indicated by an asterisk ($^*$), dictating the optimal parsimonious lag structure.}
\label{tab:var_lag_selection}
\resizebox{12.cm}{!}{
\begin{tabular}{|@{} c cccc @{}|}
\toprule
\textbf{Lag} & \textbf{AIC} & \textbf{BIC} & \textbf{FPE} & \textbf{HQIC} \\
\midrule
0  & -57.7500          & -57.7400$^*$ & 8.294e-26          & -57.7500$^*$ \\
1  & -57.7700$^*$      & -57.6700          & 8.128e-26$^*$      & -57.7400          \\
2  & -57.7600          & -57.5700          & 8.266e-26          & -57.6900          \\
3  & -57.7400          & -57.4800          & 8.371e-26          & -57.6500          \\
4  & -57.7200          & -57.3700          & 8.530e-26          & -57.6000          \\
5  & -57.7100          & -57.2800          & 8.617e-26          & -57.5600          \\
6  & -57.7000          & -57.1800          & 8.775e-26          & -57.5100          \\
7  & -57.6900          & -57.0800          & 8.866e-26          & -57.4700          \\
8  & -57.6800          & -56.9900          & 8.925e-26          & -57.4300          \\
9  & -57.6700          & -56.9000          & 9.024e-26          & -57.3900          \\
10 & -57.6500          & -56.8000          & 9.146e-26          & -57.3400          \\
\bottomrule
\multicolumn{5}{l}{\footnotesize \textit{Note:} AIC: Akaike, BIC: Schwarz Bayesian, FPE: Final Prediction Error, HQIC: Hannan-Quinn.}
\end{tabular}
}
\end{table}

\begin{table}[!h]
\centering
\caption{Contemporaneous correlation matrix of residuals obtained from the estimated VAR(1) model. The lower triangular presentation highlights immediate cross-asset shock dependencies.}
\label{tab:residual_correlation}
\resizebox{10.cm}{!}{
\begin{tabular}{@{} l cccccc @{}}
\toprule
\textbf{Asset} & \textbf{ICLN} & \textbf{QQQ} & \textbf{SPY} & \textbf{TAN} & \textbf{XLE} & \textbf{XLU} \\
\midrule
\textbf{ICLN} & 1.0000 &        &        &        &        &        \\
\textbf{QQQ}  & 0.6669 & 1.0000 &        &        &        &        \\
\textbf{SPY}  & 0.7270 & 0.9270 & 1.0000 &        &        &        \\
\textbf{TAN}  & 0.8138 & 0.5882 & 0.6163 & 1.0000 &        &        \\
\textbf{XLE}  & 0.6388 & 0.6642 & 0.7946 & 0.5519 & 1.0000 &        \\
\textbf{XLU}  & 0.3974 & 0.4060 & 0.5284 & 0.2760 & 0.3913 & 1.0000 \\
\bottomrule
\end{tabular}
}
\end{table}

\begin{table}[!h]
\centering
\caption{Brock--Dechert--Scheinkman (BDS) tests applied to the residuals of the estimated VAR(1) model. The null hypothesis states that the residual series are independently and identically distributed (i.i.d.).}
\label{tab:bds_test_results}
\resizebox{11.cm}{!}{
\begin{tabular}{@{} l ccccc @{}}
\toprule
 & \multicolumn{5}{c}{\textbf{Embedding Dimension ($m$)}} \\
\cmidrule(l){2-6}
\textbf{Asset} & \textbf{$m=2$} & \textbf{$m=3$} & \textbf{$m=4$} & \textbf{$m=5$} & \textbf{$m=6$} \\
\midrule
\textbf{ICLN} & 1.1148          & 1.7605$^{*}$    & 2.5537$^{**}$   & 3.4084$^{***}$  & 4.2830$^{***}$  \\
\textbf{QQQ}  & 1.6017          & 2.6004$^{***}$  & 3.8715$^{***}$  & 5.4732$^{***}$  & 7.3842$^{***}$  \\
\textbf{SPY}  & 1.8523$^{*}$    & 3.2842$^{***}$  & 4.9962$^{***}$  & 6.9524$^{***}$  & 9.3550$^{***}$  \\
\textbf{TAN}  & 1.0447          & 1.8050$^{*}$    & 2.8123$^{**}$   & 3.8868$^{***}$  & 5.1264$^{***}$  \\
\textbf{XLE}  & 0.0958          & 1.6283          & 2.3457$^{*}$    & 3.1661$^{**}$   & 4.0048$^{***}$  \\
\textbf{XLU}  & 0.5432          & 0.8837          & 1.1498          & 1.4111          & 1.8089$^{*}$    \\
\bottomrule
\multicolumn{6}{l}{\footnotesize \textit{Note:} $^{*}$ ($p < 0.10$), $^{**}$ ($p < 0.05$), and $^{***}$ ($p < 0.01$).}
\end{tabular}
}
\end{table}
Importantly, these results should not be interpreted as structural causal spillovers. Rather, they indicate that substantial multivariate dependence remains present across transition-related assets.

\subsection{Residual Nonlinear Dependence}

Although the VAR specification captures an important fraction of the linear dependence structure, nonlinear residual dynamics may still remain. To investigate this issue, we apply the Brock--Dechert--Scheinkman (BDS) test to the residuals of the estimated VAR(1) model.

The BDS statistics reject the null hypothesis of independently and identically distributed residuals across several embedding dimensions, particularly for SPY, QQQ, TAN, and ICLN. These results indicate that nonlinear dependence remains even after removing the linear multivariate structure through the VAR specification.

Importantly, the strongest nonlinear signatures appear in renewable-energy and growth-related assets, which are also more sensitive to innovation cycles, financing conditions, and macro-financial uncertainty.

Taken together, the empirical evidence indicates that transition-energy financial markets simultaneously exhibit:
\begin{itemize}
    \item heavy-tailed return distributions,
    \item persistent volatility dynamics,
    \item substantial multivariate dependence,
    \item and residual nonlinear predictability.
\end{itemize}

These stylized facts directly motivate the hybrid econometric architecture introduced in the next section. The Student-\textit{t} VAR component is designed to capture heavy-tailed multivariate dependence, while the recurrent residual-learning component models the remaining nonlinear dependence structure left unexplained by the linear econometric specification.

\section{Hybrid Heavy-Tailed Forecasting Framework}

This section develops the forecasting framework used to model transition-related financial returns. The methodology is explicitly designed to address the empirical regularities documented in Section~\ref{sec4}, namely:
\begin{enumerate}
    \item heavy-tailed return innovations,
    \item persistent cross-market dependence,
    \item and residual nonlinear predictability.
\end{enumerate}
The proposed framework does not treat machine learning as a substitute for econometric modeling. Instead, the econometric specification constitutes the primary modeling layer, while machine learning is introduced only as a residual correction mechanism. This distinction is important both statistically and economically. The econometric component captures the dominant linear dependence structure and heavy-tailed innovations across transition-related assets, whereas the nonlinear learner approximates the remaining predictable structure left unexplained by the multivariate Student-$t$ VAR system.

Methodologically, the framework follows the hybrid forecasting philosophy introduced by \citet{zhang2003time} and subsequently extended in econometric-machine learning hybridization studies such as \citet{pokou2024hybridization}. The central idea underlying this literature is that complex financial time series often contain both:
\begin{enumerate}
    \item a dominant linear dependence structure,
    \item and a lower-dimensional nonlinear predictable component.
\end{enumerate}
Under this decomposition, econometric models are used to extract the primary linear dynamics, while machine learning methods are applied only to the residual process. Consequently, the nonlinear learner does not replace the econometric structure but instead approximates the remaining nonlinear dependence that persists after linear filtering.
This distinction is particularly relevant for transition-energy financial markets. The empirical evidence reported in Section~\ref{sec4} indicates that:
\begin{itemize}
    \item return distributions are strongly non-Gaussian,
    \item volatility dynamics are persistent,
    \item and nonlinear dependence remains present even after linear multivariate filtering.
\end{itemize}
Accordingly, the methodology combines:
\begin{enumerate}
    \item a multivariate Student-$t$ VAR specification capturing heavy-tailed linear dependence;
    \item a nonlinear residual-learning layer approximating the remaining predictable residual structure.
\end{enumerate}
Importantly, the paper does not claim that all transition-related financial dynamics are fundamentally nonlinear or that recurrent neural architectures provide structural economic interpretation. Rather, the objective is narrower and empirically testable: to determine whether nonlinear predictability remains after controlling for heavy-tailed multivariate linear dependence.

\subsection{Multivariate Return System}

Let
\begin{equation}
y_t =
\begin{pmatrix}
r_{1t} & r_{2t} & \cdots & r_{Kt}
\end{pmatrix}^{\prime}
\in \mathbb{R}^{K}
\end{equation}
denote the vector of financial returns observed at time $t$, where $K=6$ in the empirical application. The vector includes broad-market, technology, fossil-energy, renewable-energy, solar-energy, and utility-sector ETFs.

The natural filtration generated by the information set is denoted by
\begin{equation}
\mathcal{F}_{t-1}= \sigma(y_{t-1},y_{t-2},\ldots).
\end{equation}
The process $\{y_t\}_{t\geq1}$ is assumed covariance-stationary unless otherwise specified. Following \citet{kilian2006new}, covariance stationarity requires all eigenvalues of the companion matrix to lie strictly inside the unit circle. This condition is empirically verified in the preliminary diagnostics.

\subsection{Gaussian VAR Benchmark}

We first consider the conventional Gaussian Vector Autoregressive model introduced by \citet{sims1980macroeconomics}. The VAR($p$) specification is given by
\begin{equation}
y_t = c + \sum_{i=1}^{p}A_i y_{t-i} + \varepsilon_t,
\label{eq:gaussian_var}
\end{equation}
where
$$
c\in\mathbb{R}^{K}, \quad A_i\in\mathbb{R}^{K\times K}, \quad \varepsilon_t\in\mathbb{R}^{K}.
$$
Under the Gaussian benchmark,
\begin{equation}
\varepsilon_t \mid \mathcal{F}_{t-1} \sim \mathcal{N}(0,\Sigma),
\end{equation}
with covariance matrix
$$
\Sigma \succ 0.
$$
The conditional expectation is therefore
\begin{equation}
\mathbb{E}(y_t\mid\mathcal{F}_{t-1}) = c + \sum_{i=1}^{p}A_i y_{t-i}.
\end{equation}

The Gaussian VAR provides an important benchmark because it captures multivariate lagged dependence through the autoregressive coefficient matrices $\{A_i\}_{i=1}^{p}$. In particular, each coefficient measures the marginal lagged dependence between transition-related assets.

However, this framework relies on two restrictive assumptions:
\begin{enumerate}
    \item Gaussian innovations,
    \item and purely linear propagation mechanisms.
\end{enumerate}
These assumptions become problematic in transition-related financial systems characterized by:
\begin{itemize}
    \item abrupt repricing episodes,
    \item volatility amplification,
    \item crisis-related tail events,
    \item and evolving dependence structures.
\end{itemize}
As documented in Section~\ref{sec4}, the empirical return distributions strongly reject Gaussianity and exhibit substantial heavy-tailed behavior. Consequently, the Gaussian VAR primarily serves as a baseline benchmark against which heavy-tailed specifications can be evaluated.

\subsection{Student-$t$ VAR Specification}
\subsubsection{Why Student-$t$ Innovations?}
The use of Student-$t$ innovations is motivated directly by the empirical properties of the data rather than by purely methodological convenience.
The QQ-plots and Jarque-Bera diagnostics reported in Section~\ref{sec4} show that transition-related financial returns exhibit:
\begin{itemize}
    \item excess kurtosis,
    \item asymmetric extreme realizations,
    \item and substantial departures from Gaussianity.
\end{itemize}
These features are economically plausible in transition-energy financial systems exposed to:
\begin{itemize}
    \item geopolitical disruptions,
    \item abrupt energy-price adjustments,
    \item inflationary tightening,
    \item climate-policy uncertainty,
    \item and rapid portfolio reallocations.
\end{itemize}
Under such environments, Gaussian innovations tend to underestimate the probability of large shocks and may generate unstable parameter estimates during crisis periods.

The Student-$t$ distribution provides a parsimonious heavy-tailed alternative widely used in financial econometrics \citep{bollerslev1987conditionally}. Importantly, the choice of Student-$t$ innovations does not imply that financial returns are exactly Student distributed. Rather, the Student-$t$ specification is adopted because:
\begin{enumerate}
    \item it accommodates heavy-tailed realizations more effectively than Gaussian disturbances;
    \item it preserves tractable multivariate likelihood estimation;
    \item and it provides robustness against influential extreme observations.
\end{enumerate}

Alternative heavy-tailed distributions could also be considered, including skewed-$t$ or generalized hyperbolic specifications. However, introducing additional asymmetry parameters substantially increases estimation complexity within multivariate systems and may reduce forecasting stability in moderate-dimensional settings. The Student-$t$ specification therefore represents a tractable compromise between empirical flexibility, robustness, and multivariate estimation feasibility.

\subsubsection{Model Specification}

The Student-$t$ VAR preserves the linear conditional mean structure:
\begin{equation}
y_t=c+\sum_{i=1}^{p}A_i y_{t-i} + \varepsilon_t,
\label{eq:vart}
\end{equation}
while replacing Gaussian innovations with a multivariate Student-$t$ distribution:
\begin{equation}
\varepsilon_t \mid \mathcal{F}_{t-1} \sim t_{\nu}(0,\Sigma),
\label{eq:studentt}
\end{equation}
where:
\begin{itemize}
    \item $\nu>2$ denotes the degrees-of-freedom parameter;
    \item $\Sigma\succ0$ is the positive-definite scale matrix.
\end{itemize}
Lower values of $\nu$ imply heavier tails and therefore larger probabilities of extreme realizations.

The multivariate Student-$t$ density is:
\begin{equation}
f(\varepsilon_t)
=\frac{\Gamma\left(\frac{\nu+K}{2}\right)}{\Gamma\left(\frac{\nu}{2}\right)
(\nu\pi)^{K/2}|\Sigma|^{1/2}}
\left[1+\frac{1}{\nu}\varepsilon_t^{\prime}
\Sigma^{-1} \varepsilon_t
\right]^{-\frac{\nu+K}{2}}.
\end{equation}
The corresponding log-likelihood function is:
\begin{equation}
\mathcal{L}(\Theta) = \sum_{t=p+1}^{T}
\log\, f\left(y_t-c-\sum_{i=1}^{p}A_i y_{t-i} \right),
\end{equation}
where
$$
\Theta = \{c,A_1,\ldots,A_p,\Sigma,\nu\}.
$$

\subsubsection{Scale-Mixture Interpretation}

An important property of the Student-$t$ specification is its Gaussian scale-mixture representation:
\begin{align}
\varepsilon_t \mid \tau_t
&\sim \mathcal{N}
\left(0, \frac{\Sigma}{\tau_t} \right),
\\ 
\tau_t &\sim \text{Gamma}
\left( \frac{\nu}{2}, \frac{\nu}{2} \right).
\end{align}
This representation provides an economically intuitive interpretation. During relatively stable periods, $\tau_t$ fluctuates around one and the process behaves approximately Gaussian. During stress episodes, smaller realizations of $\tau_t$ inflate conditional variance and generate heavy-tailed observations.

Consequently, the Student-$t$ specification can be interpreted as a parsimonious mechanism allowing endogenous volatility amplification during turbulent market conditions.

\subsubsection{Robustness Properties}

Another important advantage of the Student-$t$ likelihood concerns robustness to extreme observations. Unlike Gaussian maximum likelihood estimation, Student-$t$ likelihood estimation automatically downweights unusually large residuals through:
$$
\left( 1+ \frac{ \varepsilon_t^{\prime} \Sigma^{-1} \varepsilon_t}{\nu} \right)^{-1}.
$$
Consequently:
\begin{itemize}
    \item extreme observations receive lower effective influence;
    \item parameter estimates become more stable during crisis episodes;
    \item and forecasting performance becomes less sensitive to tail realizations.
\end{itemize}
This robustness property is particularly relevant in transition-related financial systems where extreme observations often correspond to economically meaningful repricing episodes rather than purely statistical anomalies.

\subsection{Residual Learning Framework}

\subsubsection{Why Residual Learning?}

The residual-learning architecture is motivated by both statistical diagnostics and economic considerations.
Importantly, the paper does not assume that nonlinear machine learning models should replace econometric structures. Instead, the nonlinear learner is introduced only after the dominant heavy-tailed multivariate dependence has already been extracted through the Student-$t$ VAR system.
This distinction is crucial because the BDS rejection documented in Section~\ref{sec4} does not, by itself, justify deep learning architectures. A rejection of the i.i.d. hypothesis merely indicates that some dependence structure remains in the residual process.

The role of residual learning is therefore more modest and more precise. The objective is to determine whether:
\begin{equation}
\mathbb{E}
\left[\varepsilon_t \mid \varepsilon_{t-1},\ldots,\varepsilon_{t-q} \right] \neq 0,
\end{equation}
even after controlling for:
\begin{itemize}
    \item heavy-tailed innovations,
    \item linear multivariate dependence,
    \item and low-order autoregressive dynamics.
\end{itemize}
Economically, this possibility is plausible in transition-related financial systems because crisis periods may generate:
\begin{itemize}
    \item delayed adjustment dynamics,
    \item heterogeneous reaction speeds across sectors,
    \item nonlinear volatility feedback,
    \item financing-condition persistence,
    \item and asynchronous repricing mechanisms.
\end{itemize}
Under such conditions, some predictable dependence may remain embedded within the residual process even after linear filtering.\\

\noindent The proposed framework therefore decomposes the return process into:
\begin{enumerate}
    \item a dominant heavy-tailed linear component;
    \item and a residual nonlinear predictable component.
\end{enumerate}
This interpretation remains substantially more conservative than claiming that transition-related financial systems are intrinsically governed by deep nonlinear mechanisms.

\subsubsection{Fundamental Hybrid Representation}

Formally, the return process is decomposed as:
\begin{equation}
y_t = \mu_t + g_t + u_t,
\label{eq:decomposition}
\end{equation}
where:
\begin{itemize}
    \item $\mu_t$ denotes the heavy-tailed linear component;
    \item $g_t$ captures residual nonlinear predictable dependence;
    \item $u_t$ represents unpredictable innovations.
\end{itemize}
The Student-$t$ VAR estimates $\mu_t$, while the nonlinear learner approximates $g_t$.\\

\noindent Let
\begin{equation}
\widehat{y}^{\,VAR-t}_{t|t-1} = \widehat{c} + \sum_{i=1}^{p} \widehat{A}_i y_{t-i}
\end{equation}
denote the Student-$t$ VAR forecast.\\

The filtered residual process is:
\begin{equation}
\widehat{\varepsilon}^{\,VAR-t}_t = y_t - \widehat{y}^{\,VAR-t}_{t|t-1}.
\label{eq:residuals}
\end{equation}
The hybrid forecast is then defined as:
\begin{equation}
\widehat{y}^{\,H}_{t|t-1} = \widehat{y}^{\,VAR-t}_{t|t-1} + f_{\theta}\left(
\widehat{\varepsilon}^{\,VAR-t}_{t-1},\ldots,\widehat{\varepsilon}^{\,VAR-t}_{t-q}\right).
\label{eq:hybrid}
\end{equation}
Equation~\eqref{eq:hybrid} constitutes the central forecasting equation of the paper. It formalizes the idea that:
\begin{enumerate}
    \item the Student-$t$ VAR extracts the dominant heavy-tailed multivariate dependence structure;
    \item the nonlinear learner approximates only the remaining predictable residual dynamics.
\end{enumerate}
This separation is essential because it preserves:
\begin{itemize}
    \item econometric interpretability,
    \item multivariate dependence structure,
    \item and statistical robustness,
\end{itemize}
while allowing flexible nonlinear residual correction.

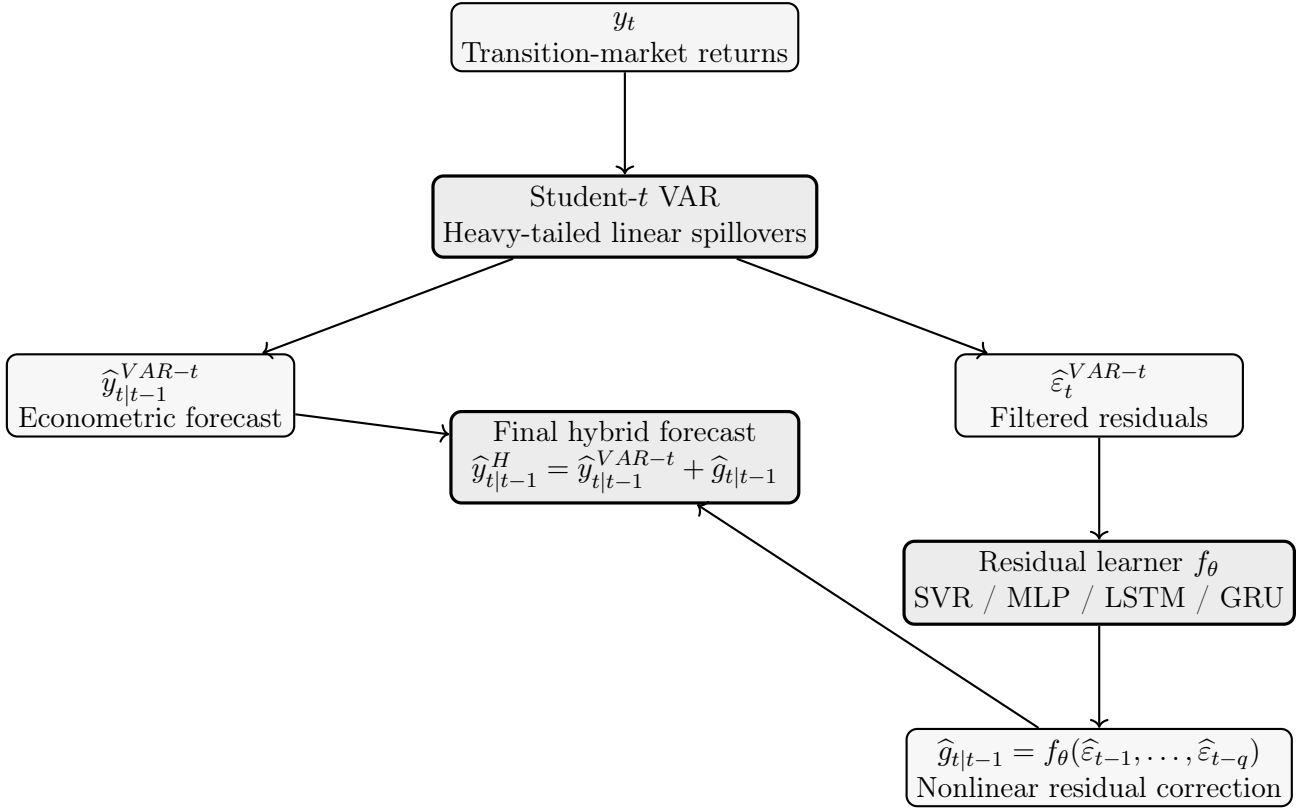
\begin{figure}[htbp]
\centering
\begin{tikzpicture}[
    node distance=1.35cm and 1.8cm,
    box/.style={
        rectangle,
        rounded corners,
        draw=black,
        thick,
        minimum width=4.0cm,
        minimum height=0.9cm,
        align=center,
        fill=gray!8
    },
    main/.style={
        rectangle,
        rounded corners,
        draw=black,
        very thick,
        minimum width=4.6cm,
        minimum height=1.0cm,
        align=center,
        fill=gray!15
    },
    smallbox/.style={
        rectangle,
        rounded corners,
        draw=black,
        thick,
        minimum width=3.8cm,
        minimum height=0.85cm,
        align=center,
        fill=gray!6
    },
    arrow/.style={->, thick}
]

% Top layer
\node[box] (data) {$y_t$\\ Transition-market returns};

% VAR-t layer
\node[main, below=of data] (vart)
{Student-$t$ VAR\\ Heavy-tailed linear spillovers};

% Split layer
\node[smallbox, below left=1.25cm and 1.8cm of vart] (forecast)
{$\widehat{y}^{\,VAR-t}_{t|t-1}$\\ Econometric forecast};

\node[smallbox, below right=1.25cm and 1.8cm of vart] (residuals)
{$\widehat{\varepsilon}^{\,VAR-t}_{t}$\\ Filtered residuals};

% Residual learning layer
\node[main, below=1.35cm of residuals] (ml)
{Residual learner $f_{\theta}$\\ SVR / MLP / LSTM / GRU};

\node[smallbox, below=of ml] (corr)
{$\widehat{g}_{t|t-1}
=
f_{\theta}
(\widehat{\varepsilon}_{t-1},\ldots,\widehat{\varepsilon}_{t-q})$\\
Nonlinear residual correction};

% Final layer
\node[main, below=2.0cm of vart] (final)
{Final hybrid forecast\\
$\widehat{y}^{\,H}_{t|t-1}
=
\widehat{y}^{\,VAR-t}_{t|t-1}
+
\widehat{g}_{t|t-1}$};

% Arrows
\draw[arrow] (data) -- (vart);

\draw[arrow] (vart) -- (forecast);
\draw[arrow] (vart) -- (residuals);

\draw[arrow] (residuals) -- (ml);
\draw[arrow] (ml) -- (corr);

\draw[arrow] (forecast) -- (final);
\draw[arrow] (corr) -- (final);

\end{tikzpicture}
\caption{Hybrid Student-$t$ VAR residual-learning architecture. The econometric layer extracts heavy-tailed linear spillovers, while the machine learning layer estimates the nonlinear predictable structure remaining in the residual process. The final forecast combines the Student-$t$ VAR prediction with the nonlinear residual correction.}
\label{fig:hybrid_architecture}
\end{figure}

\noindent Figure~\ref{fig:hybrid_architecture} summarizes the architecture of the proposed framework.

\subsubsection{Residual Learning Objective}

The nonlinear learner is estimated through:
\begin{equation}
\widehat{\theta}
=
\arg\min_{\theta}
\frac{1}{T-q}
\sum_{t=q+1}^{T}
\left\|
\widehat{\varepsilon}^{\,VAR-t}_t
-
f_{\theta}
\left(
\widehat{\varepsilon}^{\,VAR-t}_{t-1},
\ldots,
\widehat{\varepsilon}^{\,VAR-t}_{t-q}
\right)
\right\|_2^2.
\label{eq:loss}
\end{equation}

Accordingly,
\begin{equation}
f_{\theta}^{\ast}
\approx
\mathbb{E}
\left[
\widehat{\varepsilon}^{\,VAR-t}_t
\mid
\widehat{\varepsilon}^{\,VAR-t}_{t-1},
\ldots,
\widehat{\varepsilon}^{\,VAR-t}_{t-q}
\right].
\end{equation}

Consequently, any forecasting improvement generated by the nonlinear learner constitutes direct empirical evidence that predictable dependence remains after controlling for:
\begin{itemize}
    \item heavy-tailed innovations,
    \item linear multivariate dependence,
    \item and low-order autoregressive structure.
\end{itemize}

\subsection{Why Residual Learning Rather than Regime-Switching Models?}

An important question concerns the choice of residual-learning architectures rather than explicit regime-switching specifications such as Markov-switching VAR systems.

Regime-switching models constitute an important alternative for modeling structural instability and have been extensively used in financial econometrics. However, they impose a relatively specific form of nonlinear dynamics based on:
\begin{itemize}
    \item discrete latent states,
    \item abrupt transitions between regimes,
    \item and parametric regime-dependent dynamics.
\end{itemize}

The present paper adopts a different perspective. Rather than assuming that nonlinear predictability necessarily arises from a small number of latent regimes, the proposed framework allows for more flexible and continuously evolving residual dependence structures.

This distinction is particularly relevant in transition-energy financial markets where:
\begin{itemize}
    \item repricing dynamics may evolve gradually,
    \item transition uncertainty may fluctuate continuously,
    \item and dependence structures may change asynchronously across sectors.
\end{itemize}

Moreover, introducing regime-switching dynamics simultaneously with heavy-tailed multivariate estimation and nonlinear recurrent structures would substantially increase model complexity and estimation instability in moderate-dimensional systems.

Accordingly, the paper adopts a parsimonious sequential approach:
\begin{enumerate}
    \item heavy-tailed multivariate dependence is first filtered econometrically;
    \item residual nonlinear predictability is then approximated flexibly through machine learning.
\end{enumerate}

This strategy preserves interpretability and forecasting tractability while remaining sufficiently flexible to capture evolving nonlinear dependence.

\subsection{Nonlinear Residual Learners}

Four nonlinear learners are considered:
\begin{enumerate}
    \item Support Vector Regression (SVR),
    \item Multilayer Perceptron (MLP),
    \item Long Short-Term Memory networks (LSTM),
    \item and Gated Recurrent Units (GRU).
\end{enumerate}

The objective of these models is not structural economic interpretation, but nonlinear approximation of the residual process.

\subsubsection{Support Vector Regression}

SVR approximates the residual function through:
\begin{equation}
f_{\theta}(x_t)
=
\sum_{j=1}^{N}
(\alpha_j-\alpha_j^{\ast})
K(x_j,x_t)
+
b,
\end{equation}
where:
\begin{itemize}
    \item $K(\cdot,\cdot)$ is a kernel function;
    \item $\alpha_j,\alpha_j^{\ast}$ are support-vector coefficients.
\end{itemize}

\subsubsection{Multilayer Perceptron}

The MLP approximates nonlinear mappings through:
\begin{equation}
f_{\theta}(x_t)
=
W_2
\sigma(W_1x_t+b_1)
+
b_2.
\end{equation}

\subsubsection{LSTM and GRU Architectures}

LSTM and GRU architectures are specifically designed for sequential dependence and persistent temporal structures \citep{hochreiter1997long,cho2014learning}. Their use is motivated by the possibility that residual dependence in transition-related financial systems may evolve gradually through delayed adjustment mechanisms and persistent volatility interactions.

Following \citet{hochreiter1997long}, the LSTM dynamics are:
\begin{align}
i_t &= \sigma(W_i x_t + U_i h_{t-1} + b_i), \\
f_t &= \sigma(W_f x_t + U_f h_{t-1} + b_f), \\
o_t &= \sigma(W_o x_t + U_o h_{t-1} + b_o), \\
c_t &= f_t \odot c_{t-1} + i_t \odot \widetilde{c}_t.
\end{align}

Similarly, the GRU dynamics are:
\begin{align}
z_t &= \sigma(W_z x_t + U_z h_{t-1} + b_z), \\
r_t &= \sigma(W_r x_t + U_r h_{t-1} + b_r), \\
h_t &= (1-z_t)\odot h_{t-1}+z_t\odot\widetilde{h}_t.
\end{align}

Importantly, recurrent architectures are not imposed as structural models of transition-related financial systems. Their role is limited to flexible approximation of residual sequential dependence after econometric filtering.

\subsection{Forecast Evaluation}

Forecast accuracy is evaluated through out-of-sample one-step-ahead predictions.

Let
\[
e^{(m)}_{j,t}
=
y_{j,t}
-
\widehat{y}^{(m)}_{j,t|t-1}
\]
denote the forecast error for model $m$ and asset $j$.

The Root Mean Squared Error is:
\begin{equation}
RMSE_j^{(m)}
=
\left[
\frac{1}{T_{test}}
\sum_{t=1}^{T_{test}}
(e^{(m)}_{j,t})^2
\right]^{1/2},
\end{equation}
while the Mean Absolute Error is:
\begin{equation}
MAE_j^{(m)}
=
\frac{1}{T_{test}}
\sum_{t=1}^{T_{test}}
|e^{(m)}_{j,t}|.
\end{equation}
To assess statistical significance of predictive differences, the Diebold--Mariano test \citep{diebold2002comparing} is implemented.

Let
$$
d_t = L(e_t^{(a)}) - L(e_t^{(b)}),
$$
where $L(\cdot)$ denotes the loss function.
Under squared-error loss: $L(e_t)=e_t^2$.\\

The Diebold--Mariano statistic is:
\begin{equation}
DM= \frac{
\overline{d}}{\sqrt{\widehat{\Omega}_d/T_{test}}},
\end{equation}
where $\widehat{\Omega}_d$ denotes a heteroskedasticity- and autocorrelation-consistent estimator of the long-run variance of $d_t$.
This framework allows us to determine whether hybrid Student-$t$ architectures generate statistically significant forecasting improvements relative to:
\begin{itemize}
    \item purely econometric models,
    \item purely machine-learning specifications,
    \item and Gaussian benchmark systems.
\end{itemize}

\section{Empirical Results}

This section evaluates the out-of-sample forecasting performance of the proposed hybrid VAR-$t$-LSTM framework across transition-related financial assets. The objective is not merely to report predictive improvements in terms of RMSE or MAE, but to investigate whether transition-energy financial markets exhibit residual nonlinear predictability and heavy-tailed dependence structures that remain economically and statistically relevant after controlling for multivariate linear dynamics.

Importantly, the empirical analysis is deliberately structured to avoid several common pitfalls frequently associated with hybrid econometric--machine learning forecasting studies.

First, the paper does not interpret forecasting improvements as evidence of universal nonlinear market behavior. Instead, the central question is substantially narrower: whether nonlinear predictable dependence remains after controlling for heavy-tailed multivariate linear interactions.

Second, the proposed framework is not evaluated solely against weak benchmarks. The comparison includes:
\begin{itemize}
    \item standard linear VAR systems,
    \item heavy-tailed VAR-$t$ specifications,
    \item standalone machine learning models,
    \item and alternative hybrid econometric--machine learning architectures.
\end{itemize}

Third, the empirical analysis focuses explicitly on regime-sensitive forecasting behavior rather than unconditional average performance alone. This distinction is particularly important in transition-related financial systems, where forecasting complexity may change substantially during periods of macroeconomic and geopolitical stress.

Finally, the objective of the empirical section is not to claim structural causal interpretation of machine-learning dynamics. The residual learners are introduced exclusively as nonlinear approximation devices applied after econometric filtering. Consequently, any predictive gains should be interpreted as evidence of remaining nonlinear dependence rather than proof of deep structural market nonlinearities.

\subsection{Out-of-Sample Evaluation Protocol}

The forecasting exercise is conducted under a strictly recursive out-of-sample evaluation framework designed to mitigate look-ahead bias and overfitting concerns.

Let $T_{train}$ denote the initial estimation window and $T_{test}$ the out-of-sample evaluation period. At each forecasting date:
\begin{enumerate}
    \item the econometric model is re-estimated using only information available up to time $t-1$;
    \item residuals are reconstructed recursively;
    \item the nonlinear learner is trained exclusively on historical residual observations;
    \item a one-step-ahead forecast is generated for time $t$.
\end{enumerate}

Accordingly, no future information is used during parameter estimation or hyperparameter selection. To ensure full computational transparency and empirical reproducibility, Table \ref{tab:reproducibility_framework} provides a comprehensive summary of the complete implementation pipeline, including econometric estimation pro-
cedures, residual-learning architectures, hyperparameter optimization protocols, rolling-window fore-
casting design, and stochastic reproducibility controls.\\

The forecasting design therefore follows a recursive walk-forward evaluation procedure rather than a static train-test split. This distinction is important because static splits may artificially inflate forecasting performance in financial environments characterized by structural instability and regime changes.
The recursive procedure is particularly relevant for transition-energy financial markets, where:
\begin{itemize}
    \item volatility regimes evolve over time,
    \item dependence structures change across crises,
    \item and abrupt repricing episodes generate nonstationary forecasting environments.
\end{itemize}

To further reduce overfitting risk:
\begin{itemize}
    \item all machine-learning architectures use identical lag structures;
    \item hyperparameter spaces remain intentionally parsimonious;
    \item recurrent architectures are trained with early stopping procedures;
    \item and forecasting comparisons are evaluated exclusively out-of-sample.
\end{itemize}

Importantly, the dimensionality of the system remains moderate ($K=6$ assets), implying that the recurrent architectures operate within relatively low-dimensional forecasting environments. Consequently, the paper does not argue that deep learning becomes necessary because of dimensional complexity alone. Rather, the motivation for recurrent architectures arises from the possibility of persistent nonlinear sequential dependence after heavy-tailed multivariate filtering.

The proposed framework therefore follows the hybrid forecasting philosophy introduced by \citet{zhang2003time} and further extended in recent econometric-machine learning hybridization studies such as \citet{pokou2024hybridization}, whereby the econometric model captures the primary linear dependence structure while machine learning methods approximate the remaining 
nonlinear residual dynamics.

\subsection{Forecasting Performance}

Table~\ref{tab:predictive_metrics_comparison} reports the out-of-sample forecasting accuracy of all competing frameworks using RMSE and MAE metrics across the six transition-related asset classes.

\begin{table}[!h]
\centering
\caption{Out-of-sample predictive performance comparison across asset return distributions.}
\label{tab:predictive_metrics_comparison}
\setlength{\tabcolsep}{2.5pt}
\scriptsize

\begin{subtable}[b]{0.49\textwidth}
\centering
\caption{Root Mean Squared Error (RMSE)}
\label{tab:metrics_rmse}
\resizebox{8.cm}{!}{
\begin{tabular}{@{} l cccccc @{}}
\toprule
\textbf{Model} & \textbf{ICLN} & \textbf{QQQ} & \textbf{SPY} & \textbf{TAN} & \textbf{XLE} & \textbf{XLU} \\
\midrule
GRU        & 0.0204 & 0.0157 & 0.0134 & 0.0268 & 0.0235 & 0.0133 \\
LSTM       & 0.0209 & 0.0163 & 0.0137 & 0.0277 & 0.0236 & 0.0146 \\
MLP        & 0.0292 & 0.0248 & 0.0262 & 0.0330 & 0.0315 & 0.0263 \\
SVR        & 0.0213 & 0.0166 & 0.0142 & 0.0283 & 0.0243 & 0.0149 \\
VAR        & 0.0223 & 0.0172 & 0.0144 & 0.0297 & 0.0255 & 0.0159 \\
VAR-GRU    & 0.0203 & 0.0162 & 0.0138 & 0.0272 & 0.0240 & 0.0149 \\
VAR-LSTM   & 0.0204 & 0.0154 & 0.0130 & 0.0268 & 0.0233 & 0.0135 \\
VAR-MLP    & 0.0220 & 0.0174 & 0.0149 & 0.0282 & 0.0241 & 0.0154 \\
VAR-SVR    & 0.0219 & 0.0170 & 0.0142 & 0.0294 & 0.0251 & 0.0153 \\
VAR-$t$    & 0.0219 & 0.0172 & 0.0143 & 0.0293 & 0.0252 & 0.0156 \\
VAR-$t$-GRU  & 0.0169 & 0.0136 & 0.0116 & 0.0224 & 0.0207 & 0.0129 \\
\textbf{VAR-$t$-LSTM} & \textbf{0.0148} & \textbf{0.0115} & \textbf{0.0093} & \textbf{0.0195} & \textbf{0.0167} & \textbf{0.0108} \\
VAR-$t$-MLP  & 0.0244 & 0.0184 & 0.0157 & 0.0426 & 0.0242 & 0.0173 \\
VAR-$t$-SVR  & 0.0179 & 0.0137 & 0.0114 & 0.0239 & 0.0215 & 0.0132 \\
\bottomrule
\end{tabular}
}
\end{subtable}
\hfill
\begin{subtable}[b]{0.49\textwidth}
\centering
\caption{Mean Absolute Error (MAE)}
\label{tab:metrics_mae}
\resizebox{8.cm}{!}{
\begin{tabular}{@{} l cccccc @{}}
\toprule
\textbf{Model} & \textbf{ICLN} & \textbf{QQQ} & \textbf{SPY} & \textbf{TAN} & \textbf{XLE} & \textbf{XLU} \\
\midrule
GRU        & 0.0116 & 0.0091 & 0.0070 & 0.0159 & 0.0129 & 0.0071 \\
LSTM       & 0.0126 & 0.0097 & 0.0076 & 0.0174 & 0.0137 & 0.0081 \\
MLP        & 0.0192 & 0.0165 & 0.0172 & 0.0227 & 0.0208 & 0.0176 \\
SVR        & 0.0140 & 0.0112 & 0.0091 & 0.0196 & 0.0154 & 0.0092 \\
VAR        & 0.0160 & 0.0124 & 0.0098 & 0.0220 & 0.0176 & 0.0106 \\
VAR-GRU    & 0.0130 & 0.0104 & 0.0083 & 0.0181 & 0.0147 & 0.0088 \\
VAR-LSTM   & 0.0123 & 0.0094 & 0.0075 & 0.0167 & 0.0135 & 0.0078 \\
VAR-MLP    & 0.0146 & 0.0116 & 0.0093 & 0.0188 & 0.0154 & 0.0095 \\
VAR-SVR    & 0.0152 & 0.0117 & 0.0094 & 0.0212 & 0.0167 & 0.0100 \\
VAR-$t$    & 0.0157 & 0.0122 & 0.0095 & 0.0217 & 0.0173 & 0.0103 \\
VAR-$t$-GRU  & 0.0088 & 0.0071 & 0.0056 & 0.0121 & 0.0100 & 0.0060 \\
\textbf{VAR-$t$-LSTM} & \textbf{0.0079} & \textbf{0.0061} & \textbf{0.0048} & \textbf{0.0107} & \textbf{0.0088} & \textbf{0.0054} \\
VAR-$t$-MLP  & 0.0135 & 0.0106 & 0.0085 & 0.0206 & 0.0142 & 0.0093 \\
VAR-$t$-SVR  & 0.0105 & 0.0081 & 0.0063 & 0.0147 & 0.0119 & 0.0071 \\
\bottomrule
\end{tabular}
}
\end{subtable}
\end{table}

Several important empirical findings emerge.

First, purely linear econometric specifications systematically underperform hybrid architectures across nearly all assets. The standard Gaussian VAR produces the largest forecasting errors in most cases, particularly for renewable-energy assets such as ICLN and TAN. This finding suggests that Gaussian-linear dependence structures alone remain insufficient to characterize transition-related return dynamics.

Second, introducing Student-$t$ innovations improves forecasting performance relative to the Gaussian VAR benchmark, although the gains remain moderate when heavy-tailed filtering is used in isolation. This result is economically informative because it indicates that heavy tails alone do not fully explain forecasting complexity in transition-related financial systems.

Importantly, the paper does not interpret Student-$t$ innovations as a universal representation of all heavy-tailed financial behavior. The Student-$t$ specification is adopted as a parsimonious and tractable heavy-tailed benchmark capable of improving robustness to extreme observations. Alternative asymmetric heavy-tailed specifications, including skewed-$t$ or generalized hyperbolic distributions, may potentially provide additional flexibility. However, incorporating such distributions within multivariate hybrid recursive forecasting systems substantially increases estimation complexity and may reduce numerical stability in moderate-dimensional environments.

Accordingly, the empirical results should be interpreted more cautiously: the findings indicate that allowing for heavy-tailed innovations materially improves forecasting robustness, rather than implying that Student-$t$ distributions perfectly characterize transition-related financial returns.

Third, recurrent architectures systematically outperform static machine-learning models such as MLP and SVR. This result suggests that residual dependence is not purely cross-sectional but also sequentially persistent.

However, standalone recurrent architectures remain inferior to hybrid econometric--machine learning systems. The superior performance of the VAR-$t$-LSTM framework therefore supports the interpretation that optimal forecasting performance requires jointly modeling:
\begin{enumerate}
    \item heavy-tailed innovations,
    \item multivariate linear dependence,
    \item and residual nonlinear sequential dependence.
\end{enumerate}

Importantly, the empirical evidence does not imply that transition-related financial systems are fundamentally governed by deep nonlinear mechanisms. Rather, the results indicate that nonlinear predictable dependence remains after controlling for dominant heavy-tailed multivariate dynamics.

The forecasting improvements are also economically meaningful in magnitude. Relative to the Gaussian VAR benchmark, the VAR-$t$-LSTM framework reduces average RMSE by approximately one-third across assets. The gains remain relatively homogeneous across asset classes, suggesting that the proposed architecture captures a common forecasting structure rather than isolated asset-specific effects.

At the same time, renewable-energy assets such as TAN and ICLN exhibit the largest improvements. This result is economically plausible because these assets are more sensitive to:
\begin{itemize}
    \item climate-policy revisions,
    \item financing conditions,
    \item technological expectations,
    \item and abrupt transition-related repricing episodes.
\end{itemize}
Consequently, their dynamics may naturally generate stronger nonlinear dependence and heavier tail exposure than broader market indices or regulated utility sectors.
Table~\ref{tab:framework_overall_ranking} further confirms the systematic dominance of heavy-tailed hybrid frameworks.

\begin{table}[!h]
\centering
\caption{Comprehensive framework ranking based on out-of-sample predictive performance.}
\label{tab:framework_overall_ranking}
\resizebox{12.cm}{!}{
\begin{tabular}{|@{} l ccc c @{}|}
\toprule
\textbf{Model} & \textbf{Avg. RMSE Rank} & \textbf{Avg. MAE Rank} & \textbf{Overall Score} & \textbf{Final Rank} \\
\midrule
\textbf{VAR-$t$-LSTM} & 1.00 & 1.00 & 1.00 & $\textbf{1}^{\text{st}}$ \\
\textbf{VAR-$t$-GRU}  & 2.17 & 2.00 & 2.08 & $\textbf{2}^{\text{nd}}$ \\
\textbf{VAR-$t$-SVR}  & 2.83 & 3.00 & 2.92 & $\textbf{3}^{\text{rd}}$ \\
\midrule
GRU                  & 5.00 & 4.00 & 4.50 & $4^{\text{th}}$ \\
VAR-LSTM             & 4.33 & 5.00 & 4.67 & $5^{\text{th}}$ \\
LSTM                 & 6.50 & 6.00 & 6.25 & $6^{\text{th}}$ \\
VAR-GRU              & 6.33 & 7.17 & 6.75 & $7^{\text{th}}$ \\
SVR                  & 8.50 & 9.00 & 8.75 & $8^{\text{th}}$ \\
VAR-MLP              & 10.17 & 9.50 & 9.83 & $9^{\text{th}}$ \\
VAR-SVR              & 9.50 & 11.00 & 10.25 & $10^{\text{th}}$ \\
VAR-$t$-MLP          & 12.50 & 8.33 & 10.42 & $11^{\text{th}}$ \\
VAR-$t$              & 10.50 & 12.00 & 11.25 & $12^{\text{th}}$ \\
VAR                  & 11.83 & 13.00 & 12.42 & $13^{\text{th}}$ \\
MLP                  & 13.83 & 14.00 & 13.92 & $14^{\text{th}}$ \\
\bottomrule
\end{tabular}
}
\end{table}
\noindent A particularly important observation emerges from the ranking structure itself. Forecasting performance improves progressively when moving from:
\[
\small{\text{Gaussian Linear Models}
\rightarrow
\text{Heavy-Tailed Models}
\rightarrow
\text{Hybrid Architectures}
\rightarrow
\text{Heavy-Tailed Hybrid Architectures}.
}\]
This monotonic ordering strongly supports the interpretation that transition-related financial systems exhibit:
\begin{itemize}
    \item heavy-tailed innovations,
    \item multivariate dependence,
    \item and residual nonlinear predictability.
\end{itemize}

\begin{figure}[!h]
    \centering
    \includegraphics[width=1.05\linewidth]{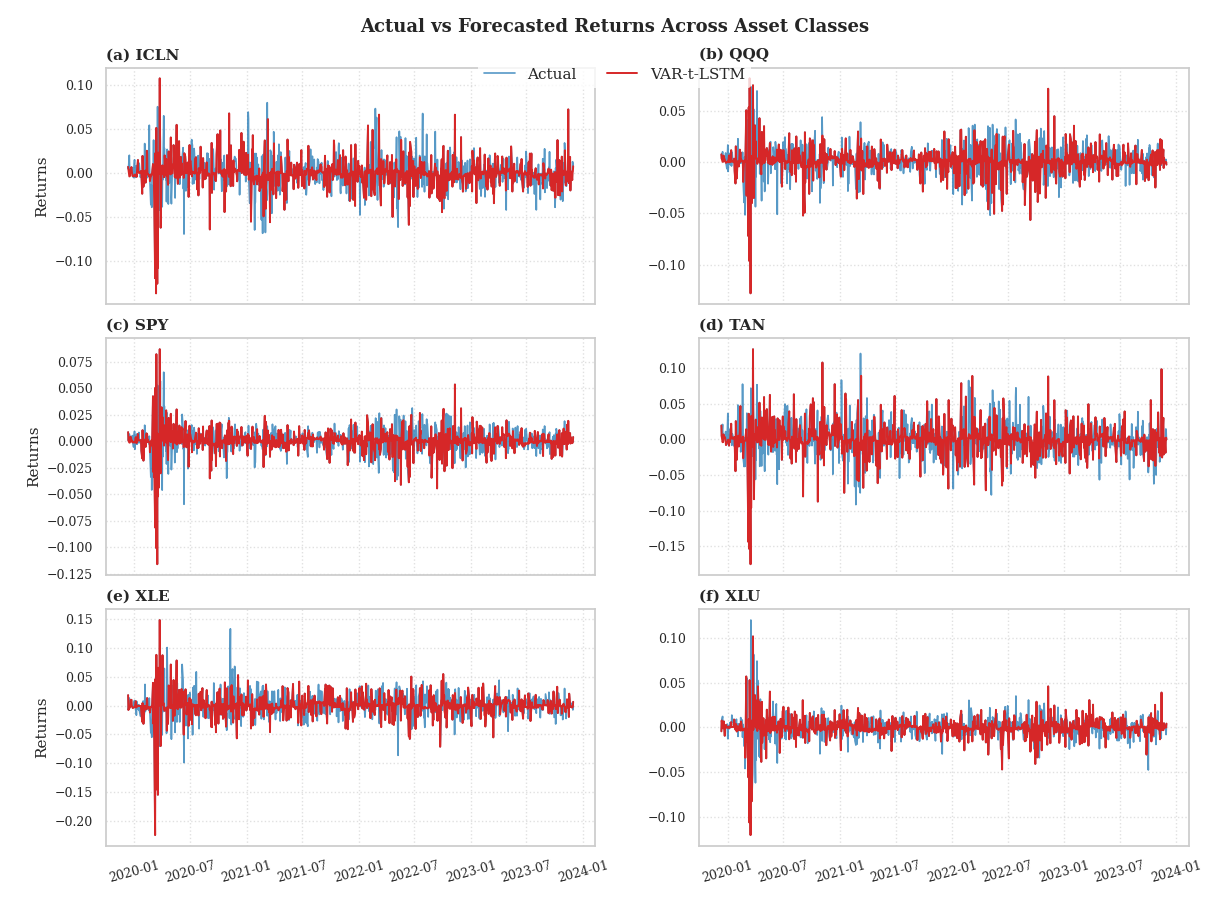}
    \caption{Time-series tracking of actual vs. predicted returns across asset classes. The figure compares realized returns with one-step-ahead forecasts generated by the proposed VAR-$t$-LSTM framework.}
    \label{fig:pred}
\end{figure}

\begin{figure}[!h]
    \centering
    \includegraphics[width=1.05\linewidth]{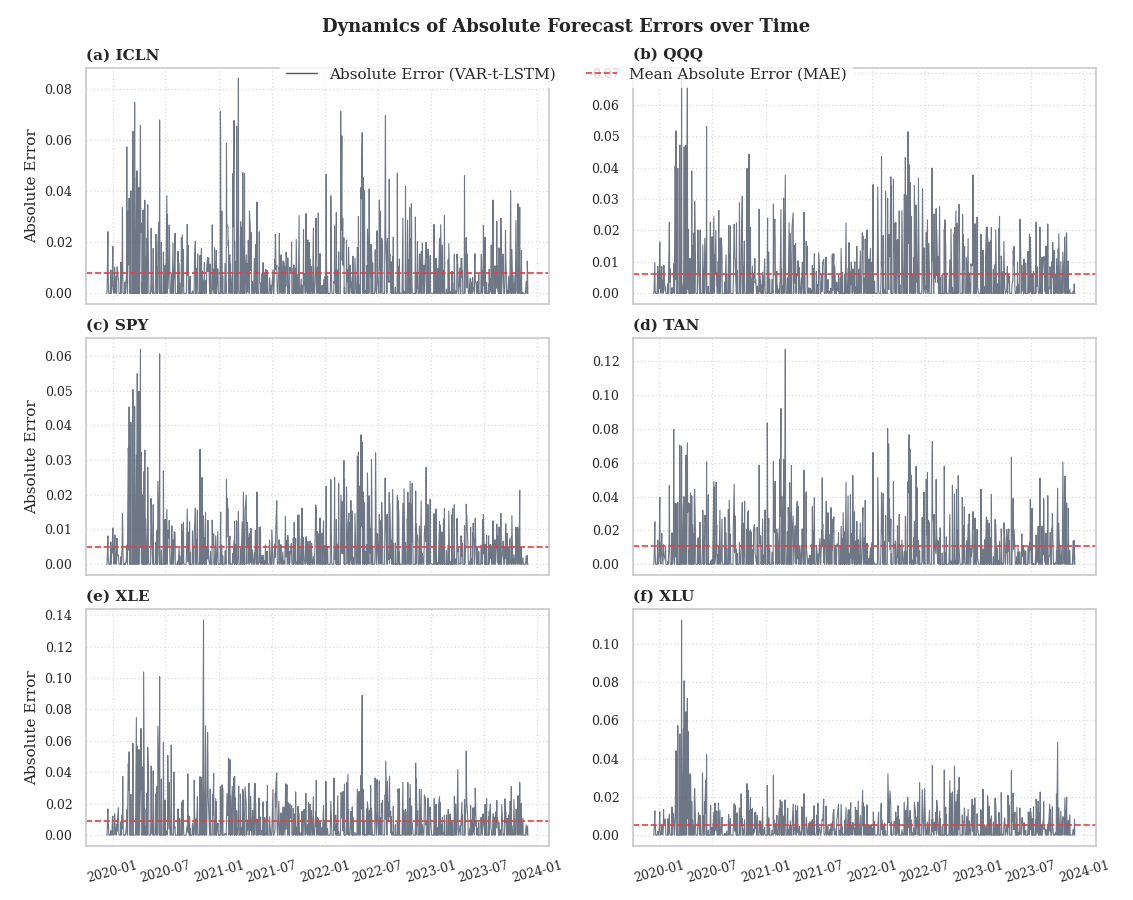}
    \caption{Temporal evolution of absolute forecasting errors across asset classes under the VAR-$t$-LSTM framework. The dashed line represents the average MAE level.}
    \label{fig:erros}
\end{figure}

\noindent Figures~\ref{fig:pred} and~\ref{fig:erros} provide complementary visual evidence.

The prediction trajectories remain closely aligned with realized returns across all assets, including periods characterized by elevated volatility. Importantly, forecasting errors are visibly heteroskedastic over time, with larger deviations concentrated around macro-financial stress episodes such as the COVID collapse and the subsequent energy-market turbulence.

This finding is economically important because it suggests that forecasting complexity itself becomes regime-dependent. In other words, transition-related financial systems appear substantially more difficult to forecast during turbulent periods characterized by synchronized repricing and elevated uncertainty.

At the same time, the average error level remains relatively stable throughout the sample, indicating that the proposed framework preserves forecasting robustness despite severe market dislocations.

Overall, the evidence reported in Tables~\ref{tab:predictive_metrics_comparison}--\ref{tab:framework_overall_ranking} and Figures~\ref{fig:pred}--\ref{fig:erros} supports a relatively cautious but economically meaningful conclusion: transition-related financial markets exhibit residual nonlinear and heavy-tailed predictability that cannot be fully captured through Gaussian-linear specifications alone.

\subsection{Diebold-Mariano Forecast Comparison Tests}

While the previous subsection documents economically meaningful forecasting improvements, statistical superiority must also be formally verified. Table~\ref{tab:diebold_mariano_results} therefore reports Diebold--Mariano (DM) tests comparing the predictive accuracy of the proposed VAR-$t$-LSTM framework against all benchmark specifications.

The results provide strong statistical evidence supporting the superiority of the proposed framework.

Against the Gaussian VAR benchmark, all DM statistics are positive and highly significant across asset classes, indicating systematic outperformance under squared-error loss. Similar conclusions hold relative to the Student-$t$ VAR specification, suggesting that heavy-tailed filtering alone remains insufficient to fully capture transition-related forecasting dynamics.

Importantly, the strongest rejection statistics are observed for renewable-energy assets such as TAN and ICLN. This finding is consistent with the interpretation that transition-sensitive sectors exhibit stronger residual nonlinear predictability and greater exposure to abrupt repricing episodes.

The proposed framework also significantly outperforms standalone machine-learning architectures. Relative to MLP and SVR models, the DM statistics are particularly large, while recurrent architectures such as LSTM and GRU constitute substantially stronger competitors.

Perhaps most importantly, the proposed framework also dominates alternative hybrid architectures. Although the VAR-$t$-GRU model performs competitively, the VAR-$t$-LSTM specification still achieves statistically significant gains across nearly all assets.

This result is economically informative. It suggests that:
\begin{enumerate}
    \item heavy-tailed econometric filtering is necessary;
    \item residual nonlinear learning is valuable;
    \item and recurrent memory structures improve forecasting stability under persistent dependence.
\end{enumerate}

At the same time, the relatively smaller DM statistics against VAR-$t$-GRU indicate that recurrent residual learning itself constitutes the dominant source of predictive improvement, while the precise recurrent architecture plays a secondary role.

Consequently, the evidence does not support the interpretation that forecasting superiority arises from arbitrary architectural complexity or overparameterization. Instead, the results are consistent with the more parsimonious interpretation that transition-related financial systems contain residual nonlinear sequential dependence after econometric filtering.

\begin{table}[!h]
\centering
\caption{Diebold--Mariano (DM) test results comparing the predictive accuracy of the proposed VAR-$t$-LSTM framework against alternative benchmarks.}
\label{tab:diebold_mariano_results}
\resizebox{12.cm}{!}{
\begin{tabular}{@{} l cccccc @{}}
\toprule
\textbf{Benchmark Model} & \textbf{ICLN} & \textbf{QQQ} & \textbf{SPY} & \textbf{TAN} & \textbf{XLE} & \textbf{XLU} \\
\midrule
\multicolumn{7}{l}{\textit{Linear \& Baseline Models}} \\
VAR         & 7.25$^{***}$ & 7.00$^{***}$ & 5.54$^{***}$ & 8.42$^{***}$ & 5.41$^{***}$ & 5.43$^{***}$ \\
VAR-$t$     & 7.31$^{***}$ & 6.82$^{***}$ & 5.38$^{***}$ & 8.42$^{***}$ & 5.19$^{***}$ & 5.27$^{***}$ \\
\midrule
\multicolumn{7}{l}{\textit{Standard Machine Learning Models}} \\
MLP         & 9.52$^{***}$ & 10.88$^{***}$ & 11.36$^{***}$ & 9.90$^{***}$ & 9.03$^{***}$ & 12.11$^{***}$ \\
SVR         & 6.15$^{***}$ & 5.86$^{***}$ & 5.20$^{***}$ & 7.10$^{***}$ & 4.42$^{***}$ & 4.21$^{***}$ \\
LSTM        & 4.98$^{***}$ & 4.57$^{***}$ & 3.73$^{***}$ & 5.63$^{***}$ & 3.54$^{***}$ & 3.31$^{***}$ \\
GRU         & 3.92$^{***}$ & 3.51$^{***}$ & 2.92$^{***}$ & 4.66$^{***}$ & 3.44$^{***}$ & 1.68$^{*}$   \\
\midrule
\multicolumn{7}{l}{\textit{Hybrid Econometric--ML Frameworks}} \\
VAR-MLP     & 3.94$^{***}$ & 3.91$^{***}$ & 3.19$^{***}$ & 4.76$^{***}$ & 3.49$^{***}$ & 2.59$^{***}$ \\
VAR-SVR     & 4.27$^{***}$ & 4.46$^{***}$ & 3.99$^{***}$ & 5.06$^{***}$ & 4.25$^{***}$ & 3.59$^{***}$ \\
VAR-LSTM    & 3.94$^{***}$ & 3.91$^{***}$ & 3.19$^{***}$ & 4.76$^{***}$ & 3.49$^{***}$ & 2.59$^{***}$ \\
VAR-GRU     & 4.27$^{***}$ & 4.46$^{***}$ & 3.99$^{***}$ & 5.06$^{***}$ & 4.25$^{***}$ & 3.59$^{***}$ \\
VAR-$t$-MLP & 5.77$^{***}$ & 6.13$^{***}$ & 5.64$^{***}$ & 5.22$^{***}$ & 4.40$^{***}$ & 5.58$^{***}$ \\
VAR-$t$-SVR & 3.19$^{***}$ & 2.66$^{***}$ & 2.43$^{**}$  & 3.66$^{***}$ & 2.74$^{***}$ & 2.58$^{***}$ \\
VAR-$t$-GRU & 1.72$^{*}$   & 2.28$^{**}$  & 2.21$^{**}$  & 1.98$^{**}$  & 2.15$^{**}$  & 2.03$^{**}$  \\
\bottomrule
\end{tabular}
}
\end{table}

\subsection{Regime-Sensitive Forecasting Performance}

A defining characteristic of transition-related financial systems concerns their sensitivity to macroeconomic and geopolitical stress regimes. Table~\ref{tab:regime_robustness_analysis} therefore evaluates predictive performance across multiple market environments, including:
\begin{itemize}
    \item the COVID crisis,
    \item the post-COVID recovery,
    \item the Ukraine-related energy shock,
    \item and the inflation-tightening regime.
\end{itemize}

The results reveal several important empirical regularities.

First, forecasting errors increase substantially during crisis periods across all models. This result confirms that forecasting complexity itself becomes state-dependent during periods of elevated uncertainty and synchronized repricing.

Second, the relative gains of the proposed framework become substantially larger during stress regimes. The strongest improvements are observed during the COVID crisis and the Ukraine-related energy shock, particularly for TAN and XLE.

This finding is economically meaningful because it suggests that residual nonlinear predictability intensifies precisely during periods characterized by:
\begin{itemize}
    \item volatility amplification,
    \item abrupt energy-market repricing,
    \item financing-condition deterioration,
    \item and elevated transition uncertainty.
\end{itemize}

Importantly, the evidence does not imply that transition-related markets are permanently nonlinear or unstable. During the post-COVID recovery regime, forecasting gains become comparatively smaller, suggesting that periods of relative stabilization partially restore linear predictability.
\newpage
\noindent Consequently, the regime analysis supports a substantially more nuanced interpretation:
\begin{quote}
\textit{transition-related financial systems exhibit state-dependent nonlinear predictability whose importance increases during periods of macro-financial stress.}
\end{quote}

This interpretation is materially more conservative and empirically defensible than claiming the existence of permanent nonlinear contagion mechanisms.

Overall, the empirical evidence reported throughout this section supports four main conclusions.
First, Gaussian-linear econometric structures remain insufficient to characterize transition-related financial returns.
Second, allowing for heavy-tailed innovations materially improves forecasting robustness.
Third, significant residual nonlinear dependence remains after econometric filtering.
Fourth, the importance of nonlinear predictability becomes substantially larger during macro-financial stress regimes.
Taken together, these findings provide the empirical motivation for the robustness exercises developed in the next section, where we further investigate whether the predictive superiority of the proposed framework remains stable across alternative market environments and stress conditions.

\section{Robustness Analysis}

This section evaluates whether the predictive superiority of the proposed VAR-$t$-LSTM framework remains stable across heterogeneous macro-financial environments, crisis episodes, and cross-sectional market structures. The objective is not merely to verify numerical stability, but to determine whether the forecasting gains documented previously reflect persistent structural features of transition-related financial systems rather than isolated sample-specific effects.

Importantly, the robustness analysis directly addresses several concerns commonly raised in forecasting studies involving hybrid econometric--machine learning architectures. In particular, the analysis investigates whether:
\begin{enumerate}
    \item predictive gains remain stable across distinct macroeconomic regimes;
    \item improvements persist across heterogeneous asset classes;
    \item forecasting superiority intensifies during stress environments;
    \item and the results are economically interpretable rather than purely statistical.
\end{enumerate}

The evidence reported below suggests that the proposed framework captures regime-sensitive nonlinear dependence structures that become increasingly relevant during periods of macro-financial instability.

At the same time, the paper deliberately avoids stronger claims regarding universal nonlinear market behavior or permanent structural instability. The empirical findings instead support a narrower and more defensible interpretation: transition-related financial systems exhibit state-dependent residual nonlinear predictability whose importance increases during turbulent market environments.

\subsection{Regime-Dependent Forecast Stability}

Table~\ref{tab:regime_robustness_analysis} reports out-of-sample forecasting performance across several macroeconomic and geopolitical regimes, including:
\begin{itemize}
    \item the COVID-19 collapse,
    \item the post-pandemic recovery,
    \item the Ukraine-related energy shock,
    \item and the inflationary monetary tightening regime.
\end{itemize}

This regime decomposition is important because unconditional forecasting averages may conceal substantial instability across heterogeneous market environments. In transition-related financial systems, dependence structures are unlikely to remain constant over time due to:
\begin{itemize}
    \item abrupt energy-price repricing,
    \item monetary-policy shifts,
    \item geopolitical fragmentation,
    \item and evolving climate-transition expectations.
\end{itemize}

Accordingly, the regime analysis evaluates whether the proposed framework remains robust precisely during periods when forecasting becomes economically most challenging.

Several important findings emerge from Table~\ref{tab:regime_robustness_analysis}.

First, all forecasting models experience substantial deterioration during crisis periods. This result confirms that forecasting complexity itself becomes strongly state-dependent during episodes of systemic uncertainty. For example, during the COVID regime, the RMSE of the benchmark VAR model increases from 0.0297 to 0.0450 for TAN and from 0.0255 to 0.0527 for XLE. Similar deteriorations are observed across all asset classes and competing specifications.

Second, the predictive superiority of the proposed framework remains stable across all market environments. In nearly every regime and for almost all assets, the VAR-$t$-LSTM specification simultaneously achieves the lowest RMSE and MAE values.

Importantly, this consistency reduces the likelihood that the reported forecasting gains merely reflect isolated crisis-specific overfitting. Instead, the results suggest that residual nonlinear dependence persists across heterogeneous market conditions.

Third, the magnitude of the forecasting gains becomes substantially larger during systemic stress periods. During the COVID regime, the proposed framework reduces the RMSE of TAN from 0.0450 under the linear VAR benchmark to 0.0240, while the corresponding reduction for XLE is from 0.0527 to 0.0284.

These improvements are economically meaningful because they occur precisely during periods characterized by:
\begin{itemize}
    \item extreme volatility amplification,
    \item abrupt energy-demand disruptions,
    \item monetary-policy interventions,
    \item and large-scale revisions in transition expectations.
\end{itemize}

At the same time, the paper does not interpret these results as evidence that transition-related markets are permanently governed by deep nonlinear dynamics. Interestingly, the forecasting gains become comparatively smaller during the post-COVID recovery regime, suggesting that partial market stabilization restores some degree of linear predictability.

Consequently, the evidence supports a substantially more nuanced interpretation:
\begin{quote}
the importance of nonlinear residual dependence increases endogenously during periods of macro-financial stress.
\end{quote}

This interpretation remains materially more conservative and empirically defensible than claiming the existence of universal nonlinear contagion mechanisms.

\begin{table}[!h]
\centering
\caption{Robustness analysis of predictive performance across distinct macroeconomic and market regimes. For each regime, $N$ denotes the number of out-of-sample trading periods. Performance is evaluated simultaneously using RMSE and MAE across all six asset classes.}
\label{tab:regime_robustness_analysis}
\scriptsize
\setlength{\tabcolsep}{3.2pt}
\renewcommand{\arraystretch}{1.1}
\resizebox{17.cm}{!}{
\begin{tabular}{@{} ll cccccc cccccc @{}}
\toprule
& & \multicolumn{2}{c}{\textbf{ICLN}} & \multicolumn{2}{c}{\textbf{QQQ}} & \multicolumn{2}{c}{\textbf{SPY}} & \multicolumn{2}{c}{\textbf{TAN}} & \multicolumn{2}{c}{\textbf{XLE}} & \multicolumn{2}{c}{\textbf{XLU}} \\
\cmidrule(lr){3-4} \cmidrule(lr){5-6} \cmidrule(lr){7-8} \cmidrule(lr){9-10} \cmidrule(lr){11-12} \cmidrule(lr){13-14}
\textbf{Regime} & \textbf{Model} & \textbf{RMSE} & \textbf{MAE} & \textbf{RMSE} & \textbf{MAE} & \textbf{RMSE} & \textbf{MAE} & \textbf{RMSE} & \textbf{MAE} & \textbf{RMSE} & \textbf{MAE} & \textbf{RMSE} & \textbf{MAE} \\
\midrule

\textbf{Full Sample} & VAR & 0.0223 & 0.0160 & 0.0172 & 0.0124 & 0.0144 & 0.0098 & 0.0297 & 0.0220 & 0.0255 & 0.0176 & 0.0159 & 0.0106 \\
($N = 999$)          & SVR & 0.0213 & 0.0140 & 0.0166 & 0.0112 & 0.0142 & 0.0091 & 0.0283 & 0.0196 & 0.0243 & 0.0154 & 0.0149 & 0.0092 \\
                     & GRU & 0.0204 & 0.0116 & 0.0157 & 0.0091 & 0.0134 & 0.0070 & 0.0268 & 0.0159 & 0.0235 & 0.0129 & 0.0133 & 0.0071 \\
                     & \textbf{VAR-$t$-LSTM} & \textbf{0.0148} & \textbf{0.0079} & \textbf{0.0115} & \textbf{0.0061} & \textbf{0.0093} & \textbf{0.0048} & \textbf{0.0195} & \textbf{0.0107} & \textbf{0.0167} & \textbf{0.0088} & \textbf{0.0108} & \textbf{0.0054} \\

\midrule
\textbf{COVID Crisis} & VAR & 0.0377 & 0.0265 & 0.0298 & 0.0204 & 0.0301 & 0.0208 & 0.0450 & 0.0327 & 0.0527 & 0.0373 & 0.0352 & 0.0245 \\
($N = 104$)           & SVR & 0.0361 & 0.0226 & 0.0293 & 0.0187 & 0.0296 & 0.0195 & 0.0428 & 0.0273 & 0.0506 & 0.0327 & 0.0337 & 0.0210 \\
                      & GRU & 0.0379 & 0.0198 & 0.0298 & 0.0160 & 0.0300 & 0.0156 & 0.0450 & 0.0242 & 0.0517 & 0.0283 & 0.0299 & 0.0156 \\
                      & \textbf{VAR-$t$-LSTM} & \textbf{0.0213} & \textbf{0.0117} & \textbf{0.0179} & \textbf{0.0093} & \textbf{0.0178} & \textbf{0.0094} & \textbf{0.0240} & \textbf{0.0136} & \textbf{0.0284} & \textbf{0.0158} & \textbf{0.0235} & \textbf{0.0119} \\

\midrule
\textbf{Post-COVID}   & VAR & 0.0210 & 0.0155 & 0.0130 & 0.0099 & 0.0093 & 0.0072 & 0.0298 & 0.0227 & 0.0216 & 0.0165 & 0.0105 & 0.0083 \\
Recovery              & SVR & 0.0204 & 0.0142 & 0.0132 & 0.0094 & 0.0100 & 0.0074 & 0.0287 & 0.0208 & 0.0204 & 0.0144 & 0.0098 & 0.0073 \\
($N = 380$)           & GRU & 0.0180 & 0.0112 & 0.0116 & 0.0072 & 0.0081 & 0.0051 & 0.0255 & 0.0162 & 0.0194 & 0.0123 & 0.0083 & 0.0055 \\
                      & \textbf{VAR-$t$-LSTM} & \textbf{0.0144} & \textbf{0.0078} & \textbf{0.0092} & \textbf{0.0050} & \textbf{0.0065} & \textbf{0.0037} & \textbf{0.0202} & \textbf{0.0112} & \textbf{0.0166} & \textbf{0.0089} & \textbf{0.0074} & \textbf{0.0043} \\

\midrule
\textbf{Ukraine Energy} & VAR & 0.0222 & 0.0173 & 0.0204 & 0.0165 & 0.0158 & 0.0125 & 0.0290 & 0.0226 & 0.0232 & 0.0185 & 0.0144 & 0.0112 \\
Shock                  & SVR & 0.0213 & 0.0151 & 0.0191 & 0.0147 & 0.0149 & 0.0108 & 0.0281 & 0.0208 & 0.0225 & 0.0172 & 0.0131 & 0.0099 \\
($N = 215$)            & GRU & 0.0197 & 0.0123 & 0.0178 & 0.0116 & 0.0138 & 0.0088 & 0.0258 & 0.0162 & 0.0198 & 0.0132 & \textbf{0.0117} & 0.0074 \\
                       & \textbf{VAR-$t$-LSTM} & \textbf{0.0169} & \textbf{0.0095} & \textbf{0.0144} & \textbf{0.0086} & \textbf{0.0111} & \textbf{0.0066} & \textbf{0.0217} & \textbf{0.0123} & \textbf{0.0161} & \textbf{0.0094} & 0.0099 & \textbf{0.0057} \\

\midrule
\textbf{Inflation}    & VAR & 0.0193 & 0.0147 & 0.0169 & 0.0132 & 0.0128 & 0.0098 & 0.0260 & 0.0199 & 0.0192 & 0.0148 & 0.0128 & 0.0099 \\
Tightening            & SVR & 0.0180 & 0.0124 & 0.0156 & 0.0113 & 0.0121 & 0.0086 & 0.0244 & 0.0175 & 0.0182 & 0.0132 & 0.0116 & 0.0084 \\
($N = 481$)           & GRU & 0.0170 & 0.0106 & 0.0146 & 0.0094 & 0.0110 & 0.0068 & 0.0231 & 0.0146 & 0.0164 & 0.0107 & 0.0108 & 0.0068 \\
                      & \textbf{VAR-$t$-LSTM} & \textbf{0.0138} & \textbf{0.0074} & \textbf{0.0117} & \textbf{0.0066} & \textbf{0.0088} & \textbf{0.0049} & \textbf{0.0184} & \textbf{0.0100} & \textbf{0.0135} & \textbf{0.0076} & \textbf{0.0089} & \textbf{0.0049} \\
\bottomrule
\end{tabular}
}
\end{table}

Table~\ref{tab:regime_rmse_improvements} complements the previous analysis by reporting the percentage RMSE reduction achieved by the proposed framework relative to the linear VAR benchmark.

\begin{table}[!h]
\centering
\caption{Proposed model's out-of-sample predictive accuracy gains over the benchmark. Values represent the percentage relative reduction in Root Mean Squared Error ($\Delta\text{RMSE}_{\%}$) achieved by the proposed hybrid architecture across distinct macroeconomic regimes and individual asset classes.}
\label{tab:regime_rmse_improvements}
\resizebox{14.cm}{!}{
\setlength{\tabcolsep}{6pt}
\renewcommand{\arraystretch}{1.15}

\begin{tabular}{@{} l cccccc | c @{}}
\toprule
\textbf{Market Regime} & \textbf{ICLN} & \textbf{QQQ} & \textbf{SPY} & \textbf{TAN} & \textbf{XLE} & \textbf{XLU} & \textbf{Average} \\
\midrule
Full Sample           & 33.59\% & 33.14\% & 35.10\% & 34.27\% & 34.55\% & 31.65\% & \textbf{33.72\%} \\
COVID Crisis          & 43.54\% & 39.91\% & 40.69\% & 46.72\% & 46.04\% & 33.32\% & \textbf{41.70\%} \\
Post-COVID Recovery   & 31.32\% & 28.81\% & 30.32\% & 32.35\% & 23.26\% & 29.02\% & \textbf{29.18\%} \\
Ukraine Energy Shock  & 23.76\% & 29.46\% & 29.60\% & 25.29\% & 30.57\% & 31.28\% & \textbf{28.33\%} \\
Inflation Tightening  & 28.50\% & 30.87\% & 30.82\% & 29.27\% & 29.83\% & 30.58\% & \textbf{29.98\%} \\
Normal Periods        & 31.24\% & 29.08\% & 30.70\% & 32.33\% & 23.58\% & 28.73\% & \textbf{29.28\%} \\
\midrule
\textbf{Average}       & \textbf{31.99\%} & \textbf{31.88\%} & \textbf{32.87\%} & \textbf{33.37\%} & \textbf{31.30\%} & \textbf{30.76\%} & \textbf{32.03\%} \\
\bottomrule
\end{tabular}
}
\end{table}

A particularly important result emerges from the temporal structure of the gains themselves. The strongest forecasting improvements occur systematically during stress environments, especially during COVID and the Ukraine-related energy shock.

This pattern is economically informative because it suggests that transition-related dependence structures become increasingly nonlinear during periods characterized by synchronized uncertainty shocks. Under such conditions:
\begin{itemize}
    \item volatility spillovers intensify,
    \item tail dependence increases,
    \item and conventional linear propagation mechanisms become less informative.
\end{itemize}

Importantly, these findings should not be interpreted as evidence that the proposed framework perfectly models crisis dynamics. Rather, the results indicate that allowing simultaneously for:
\begin{itemize}
    \item heavy-tailed innovations,
    \item multivariate dependence,
    \item and nonlinear residual propagation
\end{itemize}
substantially improves forecasting robustness during turbulent environments.

\subsection{Cross-Sectional Stability Across Asset Classes}

An important concern in hybrid forecasting studies is whether predictive improvements are driven by isolated sectors or specific assets. The evidence reported in Tables~\ref{tab:regime_robustness_analysis} and~\ref{tab:regime_rmse_improvements} strongly suggests that this is not the case here.

The forecasting gains remain remarkably homogeneous across all six ETFs despite substantial economic heterogeneity between:
\begin{itemize}
    \item broad equity markets,
    \item technology-intensive growth sectors,
    \item fossil-energy industries,
    \item renewable-energy markets,
    \item and defensive utility sectors.
\end{itemize}

This cross-sectional consistency is important because it reduces the likelihood that the predictive improvements merely reflect isolated asset-specific anomalies.

At the same time, meaningful cross-sectional asymmetries remain observable.

Renewable-energy assets such as TAN and ICLN consistently exhibit the strongest forecasting improvements during crisis regimes. By contrast, utility-sector assets such as XLU display comparatively smaller but still economically meaningful gains.

This asymmetry possesses a plausible macro-financial interpretation. Renewable-energy sectors are substantially more exposed to:
\begin{itemize}
    \item climate-policy revisions,
    \item financing-condition instability,
    \item speculative capital reallocations,
    \item and technological repricing uncertainty.
\end{itemize}

Consequently, these sectors may naturally generate stronger nonlinear residual dependence than comparatively defensive sectors such as utilities.

Importantly, the results do not imply that spillovers originate uniquely from renewable-energy markets. Rather, the evidence suggests that transition-related uncertainty propagates through multiple interconnected channels involving:
\begin{itemize}
    \item energy prices,
    \item inflation expectations,
    \item monetary-policy transmission,
    \item technology-sector valuation dynamics,
    \item and cross-sector portfolio reallocation.
\end{itemize}

Accordingly, the paper uses the term ``spillovers'' in a forecasting and dependence sense rather than as formal structural causal transmission identified through structural VAR decomposition. The objective is therefore to capture predictive cross-market dependence rather than establish causal spillover identification.

This distinction is important because it avoids overstating the interpretation of the econometric framework.

\subsection{Structural Persistence of Forecasting Gains}

\begin{figure}[!h]
    \centering
\includegraphics[width=1.05\linewidth]{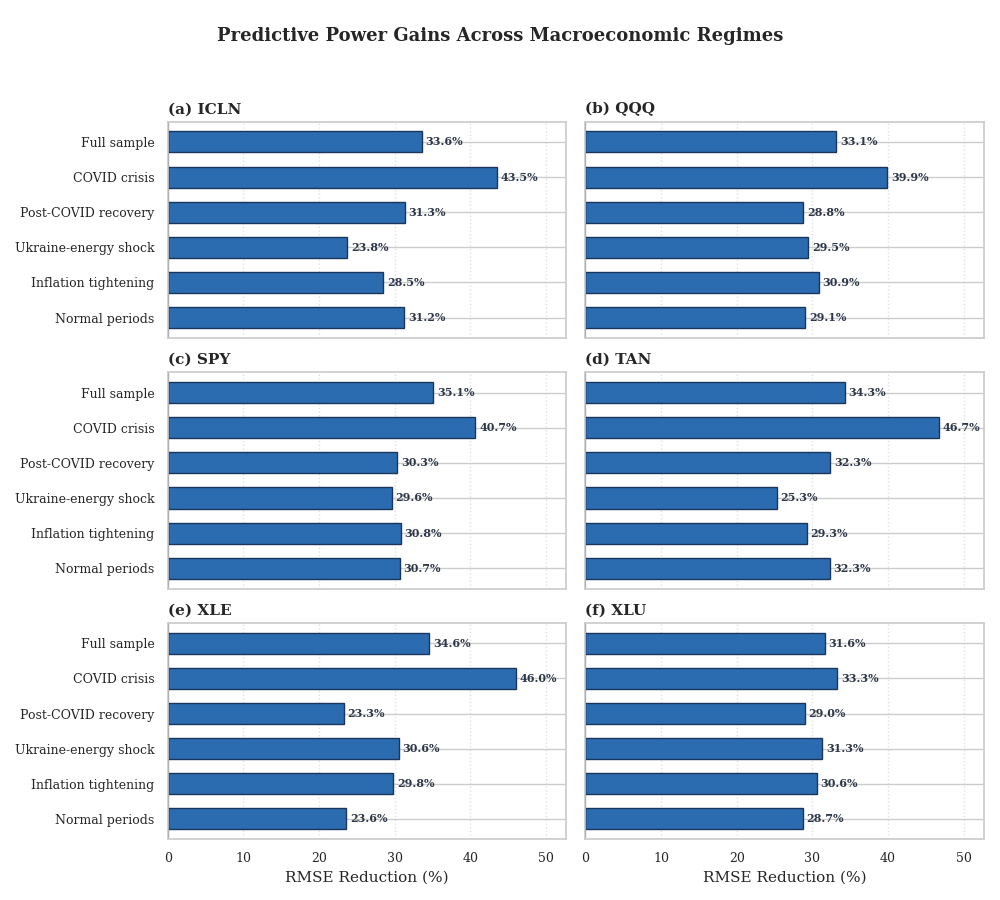}
    \caption{Predictive power enhancements across distinct market regimes and asset classes. The horizontal bar charts quantify the relative percentage decrease in forecasting RMSE compared to the linear VAR benchmark.}
    \label{fig:poower}
\end{figure}

Figure~\ref{fig:poower} provides a visual synthesis of the regime-dependent forecasting improvements.

Several important patterns emerge.

First, forecasting improvements remain positive across all regimes and all asset classes without exception. This result is particularly important because it indicates that the proposed framework never underperforms the benchmark linear specification.

Second, the gains vary systematically with market stress intensity. The strongest improvements are observed during COVID, followed by the Ukraine-related energy shock and the inflation-tightening regime.

This temporal ordering strongly supports the interpretation that nonlinear residual dependence becomes increasingly relevant during periods of elevated systemic uncertainty.

Third, the improvements are not confined exclusively to renewable-energy markets. Broad equity indices such as SPY and QQQ also experience substantial forecasting gains during stress periods, indicating that transition-related uncertainty propagates beyond the energy sector itself.

Importantly, the paper does not interpret these findings as definitive evidence of systemic causal contagion. Rather, the results suggest that macro-financial stress environments generate increasingly complex cross-market dependence structures that are insufficiently captured by linear Gaussian frameworks.

\subsection{Interpretation and Modeling Implications}

Taken together, the robustness analysis reinforces four main conclusions.

First, the forecasting superiority of the proposed framework remains stable across heterogeneous macroeconomic environments.

Second, the gains become substantially larger during stress regimes, suggesting that nonlinear residual dependence intensifies during periods of elevated uncertainty.

Third, the improvements remain cross-sectionally persistent across highly heterogeneous asset classes.

Fourth, allowing simultaneously for:
\begin{itemize}
    \item heavy-tailed innovations,
    \item multivariate econometric filtering,
    \item and residual nonlinear sequential learning
\end{itemize}
materially improves forecasting robustness in transition-related financial systems.

At the same time, the empirical evidence remains consistent with a deliberately cautious interpretation.

The paper does not claim that:
\begin{itemize}
    \item all transition-finance dynamics are fundamentally nonlinear;
    \item Student-$t$ innovations perfectly characterize financial tails;
    \item or recurrent architectures possess direct structural economic interpretation.
\end{itemize}

Instead, the results support the narrower conclusion that transition-related financial systems exhibit residual nonlinear and heavy-tailed predictability whose importance becomes increasingly pronounced during macro-financial stress periods.

These findings provide the foundation for the next section, which discusses the broader implications of regime-sensitive forecasting instability for climate finance and transition-risk modeling.

%%%
\section{Implications for Climate Finance}

The empirical evidence reported in the previous sections has implications that go beyond conventional forecasting accuracy. The results indicate that transition-energy financial markets exhibit regime-sensitive predictability: forecasting errors and model rankings change materially across macro-financial environments, and the gains of the proposed framework are largest during stress episodes. This pattern is important because the economic value of a forecasting model is highest precisely when market uncertainty, tail exposure, and portfolio rebalancing costs increase.

The interpretation developed in this section is deliberately cautious. The paper does not claim to identify structural causal spillovers in the sense of a fully specified structural VAR or network decomposition. Instead, the results show that cross-market predictive dependence becomes more complex during transition-related stress regimes, and that this complexity is better captured when heavy-tailed econometric filtering is combined with nonlinear residual learning. This distinction is important because it prevents overstating the empirical contribution while clarifying its relevance for climate-finance applications.

\subsection{Stress Forecasting and Economic Significance}

A central implication concerns stress forecasting. In climate-finance applications, average forecasting accuracy over the full sample is informative but insufficient. Investors, risk managers, and supervisors are primarily concerned with model performance during periods of abrupt repricing, volatility amplification, and tail-risk synchronization.

The regime evidence reported above shows that the VAR-$t$-LSTM framework generates its largest gains during stressed environments, especially during the COVID crisis and the Ukraine-related energy shock. This result gives the forecasting improvements an economic interpretation. The model is not only reducing unconditional prediction errors; it is improving forecasts precisely when transition-related assets become harder to predict and when forecasting errors are most costly.

This finding matters for climate-finance modeling because transition risk is unlikely to materialize as a smooth and gradual adjustment process only. Episodes of energy-market disruption, inflationary tightening, geopolitical stress, or abrupt policy reassessment can generate discontinuous repricing across fossil-energy, renewable-energy, technology, and utility sectors. Under such conditions, models based on stable Gaussian-linear dependence may underestimate the magnitude and persistence of forecast errors.

The evidence therefore supports the use of stress-sensitive forecasting frameworks in transition-risk analysis. The economic relevance of the proposed framework comes from its ability to reduce forecasting errors during high-volatility regimes, not merely from improving average RMSE or MAE values over calm periods.

\subsection{Portfolio Allocation under Transition Risk}

The results also have implications for portfolio allocation. Standard portfolio models typically rely on estimated covariance matrices and assume relatively stable dependence structures across assets. In transition-energy markets, this assumption may be restrictive because the relationship between fossil-energy, renewable-energy, technology, and utility assets changes across macro-financial regimes.

The empirical findings suggest that transition-related portfolio risk is not only a matter of higher volatility. It is also associated with instability in predictive dependence. During stress periods, assets that may appear partially diversified in normal times can become jointly exposed to common macro-financial shocks such as inflation, energy-price disruptions, and financing-condition tightening.

This has direct implications for climate-sensitive portfolios. Renewable-energy assets, for instance, are often treated as long-run beneficiaries of decarbonization. However, the empirical evidence shows that they remain highly exposed to short-run financial conditions, especially interest-rate expectations, technology-sector repricing, and speculative capital flows. Similarly, fossil-energy assets remain strongly affected by geopolitical shocks and energy-demand uncertainty. The simultaneous repricing of these sectors during crises reduces the reliability of static diversification rules.

From an allocation perspective, the results suggest that portfolio models should incorporate time-varying and stress-sensitive forecasting information. Forecast improvements during crisis regimes may translate into better risk control, more timely rebalancing, and improved downside-risk management. While the paper does not estimate a full utility-based allocation exercise, the documented reduction in forecast errors during stress periods provides evidence of economic significance for investors exposed to transition-related assets.

\subsection{Tail-Risk Monitoring and Transition-Risk Surveillance}

The findings are also relevant for transition-risk surveillance. The stylized facts show substantial departures from Gaussianity, while the forecasting results indicate that heavy-tailed specifications improve predictive robustness. However, the paper does not equate heavy tails with the Student-$t$ distribution. The Student-$t$ specification should be interpreted as a parsimonious and tractable heavy-tailed approximation rather than a complete description of financial tail behavior.

This distinction is important because transition-related financial returns may also exhibit asymmetry, skewness, and downside-tail dependence. Alternative specifications, such as skewed-$t$, generalized hyperbolic, or EVT-based models, may capture additional features of tail risk. The contribution of this paper is more specific: it shows that even a relatively parsimonious heavy-tailed multivariate filter improves forecasting performance, especially when combined with nonlinear residual learning.

For supervisory and risk-monitoring purposes, this result suggests that Gaussian-linear stress-testing tools may underestimate the severity of transition-related market adjustments. During periods of macro-financial stress, tail events may cluster across fossil-energy, renewable-energy, technology, and utility assets. A monitoring framework that ignores heavy-tailed innovations and residual nonlinear dependence may therefore understate both the frequency and the persistence of large forecast errors.

The practical implication is that climate-risk surveillance should explicitly account for tail-sensitive and regime-sensitive predictive structures. This is particularly relevant for institutions monitoring energy-sector exposures, green portfolios, and transition-sensitive equity indices.

\subsection{Interpreting Cross-Market Dependence}

A further implication concerns the interpretation of cross-market dependence. The empirical evidence shows that transition-energy assets are strongly interconnected, but the paper does not claim to provide a formal structural spillover decomposition. The term ``spillover'' should therefore be understood in a predictive and reduced-form sense: shocks and forecast errors in one segment of the transition-energy system contain information about subsequent movements in other segments.

This distinction is important for avoiding overinterpretation. The VAR and VAR-$t$ components capture multivariate predictive dependence, while the residual-learning component captures remaining nonlinear sequential dependence. These elements provide interpretable forecasting layers, but they do not identify causal transmission channels without additional structural restrictions.

Within this reduced-form interpretation, the results remain economically meaningful. Fossil-energy assets, renewable-energy ETFs, technology-heavy indices, and utilities are exposed to common macro-financial forces. Energy shocks affect inflation expectations and monetary-policy paths. Monetary tightening affects long-duration renewable and technology assets. Geopolitical disruptions affect fossil-energy profitability and utility infrastructure expectations. These channels can generate nonlinear predictive dependence even without requiring a fully identified causal network.

Thus, the paper contributes to climate finance by showing that transition-related dependence is empirically stronger and less linear during stress periods. This provides a disciplined basis for future work combining heavy-tailed forecasting, structural connectedness analysis, and portfolio-based climate stress testing.

\subsection{Implications for Climate-Finance Regulation}

The results have implications for regulatory approaches to climate-related financial risk. Many climate stress-testing exercises rely on scenario analysis with relatively smooth transmission mechanisms. The evidence in this paper suggests that such representations may be incomplete when financial markets respond abruptly to transition-related shocks.

During stress regimes, the predictive gains of the proposed framework increase materially. This indicates that market dynamics become more nonlinear and tail-sensitive precisely when supervisory attention is most needed. Regulatory frameworks that rely exclusively on linear sensitivity analysis may therefore underestimate the instability of transition-related asset prices.

The findings support the development of dynamic climate-risk monitoring systems that combine three elements: heavy-tailed risk measurement, multivariate dependence modeling, and stress-regime forecasting. Such systems would not replace structural climate scenarios, but they could complement them by identifying periods in which market-based transition risk becomes unusually difficult to forecast.

Overall, the empirical results suggest that transition risk should not be viewed only as a slow-moving repricing process. It can also emerge through abrupt, stress-dependent market adjustments in which tail events, volatility persistence, and nonlinear predictive dependence interact. This perspective provides the basis for the concluding discussion.

\section{Conclusion}

This paper examined the predictive dynamics of transition-related financial markets through a hybrid forecasting framework combining multivariate Student-$t$ Vector Autoregressions with nonlinear residual-learning architectures.

The analysis was motivated by a relatively specific empirical question rather than by a purely algorithmic objective. The paper investigated whether transition-related financial returns contain residual nonlinear predictability after controlling for heavy-tailed multivariate linear dependence. In this respect, the contribution should not be interpreted as the proposal of a generic deep-learning forecasting architecture, but rather as an empirical assessment of the statistical structure underlying transition-energy financial systems.

Several findings emerge consistently throughout the analysis.

First, the empirical evidence confirms that Gaussian-linear specifications remain insufficient to characterize transition-related asset returns. Across all asset classes, the descriptive and econometric diagnostics revealed substantial departures from Gaussianity, persistent volatility clustering, and remaining nonlinear dependence after linear multivariate filtering. These features were particularly pronounced for renewable-energy and fossil-energy assets during stress periods.

Second, allowing for heavy-tailed innovations materially improved forecasting robustness. However, the paper deliberately avoids equating heavy tails with the Student-$t$ distribution itself. The Student-$t$ specification was adopted as a parsimonious and tractable heavy-tailed approximation rather than as a definitive characterization of financial tail behavior. Transition-related returns may also exhibit asymmetry and more complex tail structures that could potentially be captured through alternative distributions such as skewed-$t$, generalized hyperbolic, or EVT-based specifications. The empirical contribution of the paper is therefore narrower and more defensible: even a relatively parsimonious heavy-tailed multivariate specification substantially improves forecasting performance relative to Gaussian benchmarks.

Third, the results indicate that residual nonlinear dependence remains economically and statistically relevant after econometric filtering. The recurrent residual-learning architectures systematically outperformed both purely econometric specifications and standalone machine-learning models. At the same time, the evidence does not support the interpretation that transition-finance systems are fundamentally governed by opaque deep nonlinear mechanisms. Rather, the forecasting gains suggest that some predictable sequential dependence persists beyond what can be captured through heavy-tailed linear dynamics alone.

An important result concerns the regime dependence of the forecasting gains. The strongest improvements were consistently observed during periods of macro-financial stress, especially during the COVID collapse and the Ukraine-related energy shock. This finding is economically meaningful because forecasting accuracy becomes most valuable precisely during periods characterized by abrupt repricing, elevated volatility, and synchronized uncertainty shocks.

Accordingly, the paper contributes less through unconditional forecasting superiority alone than through its ability to improve predictive robustness under stressed market conditions. In this sense, the results possess a clearer economic interpretation than standard forecasting exercises based solely on average RMSE reductions over stable periods.

The evidence also suggests that transition-related financial instability extends beyond renewable-energy assets themselves. Significant forecasting gains were observed simultaneously for broad equity indices, technology-intensive sectors, fossil-energy markets, and utility-sector assets. This pattern indicates that transition-related uncertainty propagates through a broader macro-financial environment involving energy prices, inflation expectations, financing conditions, monetary-policy transmission, and sectoral capital reallocation.

Importantly, the paper does not claim to identify structural spillovers in a causal sense. The dependence structures captured by the VAR and residual-learning components should instead be interpreted as predictive cross-market linkages. This distinction is essential because the framework is designed to model forecasting dependence rather than establish structural transmission channels through fully identified causal systems.

From a methodological perspective, the results suggest that neither heavy-tailed econometric specifications nor machine-learning architectures alone are sufficient to characterize transition-related financial dynamics adequately. The empirical evidence instead supports a layered interpretation in which:
\begin{enumerate}
    \item the multivariate econometric component captures the dominant linear and heavy-tailed dependence structure;
    \item the nonlinear learner approximates the remaining predictable residual dependence;
    \item and the importance of this residual dependence becomes substantially larger during stressed market environments.
\end{enumerate}

This interpretation remains intentionally more conservative than broader claims frequently encountered in forecasting studies involving hybrid artificial-intelligence architectures.

The paper also carries implications for climate-finance modeling and transition-risk surveillance. In particular, the findings suggest that linear Gaussian frameworks may underestimate forecasting instability during episodes of macro-financial stress. Because transition-related uncertainty can generate abrupt repricing across interconnected sectors, forecasting systems capable of jointly accounting for heavy-tailed innovations and regime-sensitive dependence may provide a more realistic representation of transition-finance dynamics.

However, several limitations remain.
First, the analysis relies on reduced-form predictive dependence rather than structural identification. Future work could therefore integrate structural connectedness frameworks, network-based spillover decomposition, or regime-switching dependence systems.
Second, while the Student-$t$ specification improves robustness to extreme observations, alternative asymmetric heavy-tailed distributions may capture additional dimensions of transition-related tail risk.
Third, the paper focuses on daily ETF dynamics and does not incorporate option-implied measures, carbon-market variables, or high-frequency climate-policy uncertainty indicators, all of which could enrich the characterization of transition-related financial instability.

Future research could also investigate whether the predictive gains documented here translate into economically measurable portfolio improvements through utility-based allocation exercises, downside-risk management, or climate stress-testing applications.

Overall, the empirical evidence suggests that transition-related financial markets are characterized by a combination of heavy-tailed innovations, cross-market predictive dependence, and regime-sensitive nonlinear dynamics that become increasingly important during macro-financial stress periods. Capturing these dimensions jointly appears necessary for improving forecasting robustness in transition-finance environments without overstating the structural interpretation of machine-learning architectures.

\section*{Data Availability}

The financial data used in this study are publicly available and were collected from Yahoo Finance through the \texttt{yfinance} Python API. The computational implementation and empirical forecasting pipeline are fully reproducible from the information provided in the manuscript and supplementary methodological appendix.

\section*{Declaration of Competing Interest}

The authors declare that they have no known competing financial interests or personal relationships that could have appeared to influence the work reported in this paper.

\bibliography{name}

\setcounter{figure}{0}
\renewcommand{\thefigure}{\Alph{section}\arabic{figure}}

\setcounter{table}{0}
\renewcommand{\thetable}{\Alph{section}\arabic{table}}
%

%\newpage

\begin{appendix}
\section{Reproducibility}

%========================================================%
\begin{center}
\footnotesize
\setlength{\LTleft}{0pt}
\setlength{\LTright}{0pt}

\begin{longtable}{p{4.1cm} p{10.7cm}}

\caption{Comprehensive reproducibility and computational transparency framework of the proposed hybrid forecasting architecture. The table documents all critical implementation layers, including econometric estimation, hyperparameter optimization, deep-learning training protocols, rolling-window evaluation design, software ecosystem, and stochastic control procedures. Such extensive disclosure aims to ensure full empirical replicability and methodological transparency in accordance with emerging standards in computational finance and machine learning research.}
\label{tab:reproducibility_framework}
\\

\toprule
\textbf{Component} & \textbf{Implementation Details} \\
\midrule
\endfirsthead

\multicolumn{2}{c}
{{\bfseries Table \thetable\ -- continued from previous page}} \\
\toprule
\textbf{Component} & \textbf{Implementation Details} \\
\midrule
\endhead

\midrule
\multicolumn{2}{r}{{Continued on next page}} \\
\endfoot

\bottomrule
\endlastfoot

%========================================================%
\multicolumn{2}{l}{\textbf{Data and Market Universe}}\\
\midrule

Financial assets &
SPY, QQQ, XLE, ICLN, TAN, and XLU exchange-traded funds representing broad equity markets, technology growth, fossil energy, renewable energy, solar energy, and utilities infrastructure. \\

Data source &
Daily adjusted closing prices downloaded from Yahoo Finance through the \texttt{yfinance} Python API. \\

Sample period &
January 2010 -- November 2023. \\

Return construction &
Logarithmic returns computed as
$
r_t = \log(P_t) - \log(P_{t-1}).
$
\\

Training/test split &
Rolling expanding-window forecasting design with an initial training window of 2501 observations and an out-of-sample evaluation period of approximately 1000 trading days. \\

Forecast horizon &
One-step-ahead daily return forecasting. \\

%========================================================%
\midrule
\multicolumn{2}{l}{\textbf{Econometric Layer}}\\
\midrule

Baseline linear model &
Vector autoregressive model:
$
y_t = c + \sum_{i=1}^{p} A_i y_{t-i} + \varepsilon_t.
$
\\

Heavy-tailed extension &
Student-$t$ VAR specification with multivariate Student innovations:
$
\varepsilon_t \sim t_\nu(0,\Sigma).
$
\\

Lag-order selection &
Optimal lag order selected through Optuna-based minimization of validation forecasting error using rolling chronological validation. \\

Cointegration analysis &
Johansen trace test implemented using \texttt{statsmodels.tsa.vector\_ar.vecm}. \\

Stationarity diagnostics &
Augmented Dickey--Fuller tests conducted for all return series. \\

Residual diagnostics &
ARCH-LM tests, Jarque--Bera tests, autocorrelation analysis, and BDS nonlinear dependence tests. \\

Covariance stabilization &
Symmetrization and eigenvalue regularization applied to covariance matrices prior to Cholesky factorization. \\

Student-$t$ estimation &
Maximum-likelihood estimation implemented using iterative optimization with adaptive covariance regularization and stabilized Cholesky decomposition. \\

Likelihood optimization &
Hybrid Adam/L-BFGS optimization framework used for multivariate Student-$t$ likelihood estimation. \\

%========================================================%
\midrule
\multicolumn{2}{l}{\textbf{Residual Learning Framework}}\\
\midrule

Hybrid forecasting equation &
\[
\widehat{y}^{H}_{t|t-1}
=
\widehat{y}^{VAR-t}_{t|t-1}
+
f_{\theta}(\widehat{\varepsilon}_{t-1},\ldots,\widehat{\varepsilon}_{t-q}).
\]
\\

Residual-learning objective &
The nonlinear learner models the predictable nonlinear structure remaining after extraction of heavy-tailed linear spillovers. \\

Residual lag structure &
Five residual lags used for all nonlinear learners unless otherwise specified. \\

Residual standardization &
Residual sequences normalized using rolling mean-standard deviation scaling prior to neural-network estimation. \\

Residual feature construction &
Supervised residual datasets constructed through lagged multivariate residual embeddings. \\

Sequential dependency modeling &
Temporal nonlinear dependencies learned through recurrent neural-network architectures operating on residual sequences. \\

%========================================================%
\midrule
\multicolumn{2}{l}{\textbf{Machine Learning Models}}\\
\midrule

SVR specification &
Radial-basis-function kernel with Optuna optimization over $C$, $\epsilon$, and kernel parameters. \\

MLP architecture &
Fully connected feedforward neural network with one to three hidden layers optimized via Optuna. \\

LSTM architecture &
Multivariate sequence-to-sequence recurrent architecture implemented in PyTorch. \\

GRU architecture &
Multivariate gated recurrent unit architecture implemented in PyTorch. \\

Sequence length &
Five-step residual sequences used as recurrent inputs. \\

Optimization algorithm &
Adam optimizer used for all neural-network training procedures. \\

Loss function &
Mean squared forecasting error. \\

Hyperparameter optimization &
Bayesian hyperparameter optimization performed with Optuna using Tree-structured Parzen Estimators (TPE). \\

Validation strategy &
Chronological train-validation splits preserving time-order dependence. \\

Update frequency &
Hybrid residual learners re-estimated every 20 rolling windows to balance adaptability and computational efficiency. \\

Early stopping strategy &
Best-performing neural-network weights retained according to validation forecasting loss. \\

Batch normalization/scaling &
Input standardization systematically applied before nonlinear estimation. \\

%========================================================%
\midrule
\multicolumn{2}{l}{\textbf{Forecast Evaluation}}\\
\midrule

Evaluation metrics &
RMSE, MAE, relative forecasting improvement, and average ranking statistics. \\

Statistical comparison &
Diebold--Mariano predictive accuracy tests conducted under squared-error loss functions. \\

Robustness analysis &
Forecast comparisons conducted across multiple benchmark econometric and machine-learning models. \\

Regime analysis &
Separate evaluations performed during COVID crisis, post-COVID recovery, Ukraine energy shock, inflation-tightening regime, and normal periods. \\

Economic interpretation &
Performance gains analyzed through the lens of nonlinear spillover amplification, systemic stress propagation, and transition-risk asymmetries. \\

Cross-sectional consistency &
Forecast improvements evaluated simultaneously across all six energy-transition asset classes. \\

Out-of-sample integrity &
Strict recursive forecasting without information leakage or look-ahead bias. \\

%========================================================%
\midrule
\multicolumn{2}{l}{\textbf{Computational Environment}}\\
\midrule

Programming language &
Python 3.12. \\

Core scientific libraries &
\texttt{NumPy}, \texttt{Pandas}, \texttt{SciPy}, \texttt{Statsmodels}, \texttt{Scikit-learn}, \texttt{PyTorch}, \texttt{Optuna}, \texttt{Matplotlib}, and \texttt{Seaborn}. \\

Deep learning backend &
PyTorch with optional CUDA GPU acceleration. \\

Random seeds &
Global reproducibility ensured through synchronized random seeds across NumPy, Python, and PyTorch environments. \\

Rolling-window protocol &
Strict chronological forecasting without future information contamination. \\

Validation design &
Time-series validation preserving temporal ordering. \\

Code structure &
Fully modular implementation separating econometric estimation, residual extraction, hyperparameter tuning, recurrent learning, and forecast evaluation pipelines. \\

Parallel computation &
GPU-compatible recurrent architectures enabling scalable deep-learning estimation. \\

Numerical stability &
Automatic covariance regularization and jitter correction implemented for matrix decompositions. \\

%========================================================%
\midrule
\multicolumn{2}{l}{\textbf{Reproducibility Standards}}\\
\midrule

Deterministic initialization &
Controlled stochastic initialization for all optimization procedures. \\

Transparent hyperparameter search &
All search spaces, optimization criteria, and training configurations explicitly disclosed. \\

Replicable forecasting pipeline &
The complete empirical workflow can be reproduced from raw data acquisition to final predictive evaluation. \\

Robustness verification &
Forecast superiority validated across multiple metrics, asset classes, and macro-financial regimes. \\

Transparency objective &
The empirical architecture is designed to satisfy emerging standards of computational reproducibility in climate finance and financial machine learning research. \\

Open methodological structure &
All estimation blocks remain modular, interpretable, and independently reproducible. \\

\end{longtable}
\end{center}
%========================================================%

\end{appendix}

\end{document}